\documentclass[review]{elsarticle}

\usepackage{hyperref}
\usepackage{amssymb,amsmath,amsthm}
\usepackage{subcaption,caption}
\usepackage{appendix}
\usepackage{color}
\usepackage{booktabs}
\usepackage{threeparttable}
\usepackage{fullpage}

\begin{document}

\begin{frontmatter}
\title{Synaptic delay induced macroscopic dynamics of the large-scale network of Izhikevich neurons 
}

\author{L.~Chen}
\ead{L477chen@uwaterloo.ca}

\author{S.~A.~Campbell\corref{mycorrespondingauthor}}
\cortext[mycorrespondingauthor]{Corresponding authors}
\ead{sacampbell@uwaterloo.ca}

\address{Department of Applied Mathematics and Centre for Theoretical Neuroscience,\\ University of Waterloo, Waterloo, ON, N2L 3G1, Canada}

\begin{abstract}

We consider a large network of Izhikevich neurons. Each neuron has a  quadratic integrate-and-fire type model with a recovery variable modelling spike frequency adaptation (SFA). We introduce a biologically motivated synaptic current expression and a delay in the synaptic transmission. Following the Ott-Antonsen theory, we reduce the network model to a mean-field system of delayed differential equations. 
Numerical bifurcation analysis allows us to locate higher-codimension bifurcations and to identify the regions in the parameter space where the network exhibits changes in the macroscopic dynamics, including transitions between states where the individual neurons exhibit asynchronous tonic firing and different types of synchronous bursting.
We investigate the impact of the heterogeneity of the quenched input current, the SFA mechanism and the synaptic delay on macroscopic dynamics. 
Essentially, all the results follow one principle, that is, the emergence of collective oscillations (COs) is due to a balance between the input currents, which cause neurons to spike, the adaptation current, which terminates spiking, and the synaptic current, which favours spiking in the excitatory network, but hinders it in the inhibitory case.
In the limit that the heterogeneity goes to zero, our perturbation and bifurcation analysis show that the behaviour of the mean-field model remains consistent, although this limit breaks an assumption of the model reduction.
For a single population of neurons with SFA, the synaptic delay has little effect on the generation of COs for weak coupling, but favours their emergence beyond that, and even induces new macroscopic dynamics. In particular,  Torus bifurcations may occur, and these are a crucial mechanism for the emergence of population bursting with two nested frequencies. We discuss how these solutions may relate to cross-frequency coupling which is potentially relevant for understanding healthy and pathological brain functions. 
%
Most parameter values in this paper are fit to hippocampal CA3 pyramidal neuron data.
We anticipate our results will provide insights into model-based inference of neurological mechanisms 
from the perspective of theoretical neuroscience.
\end{abstract}

\begin{keyword}
Large network \sep Izhikevich neuron \sep Synaptic delay \sep Mean field \sep Bifurcation analysis \sep Population bursting
\end{keyword}
\end{frontmatter}



\section{INTRODUCTION}

Many neural systems involve networks with a large number of synaptically coupled neurons. Whether these networks can perform their functions or not depends on the emergent macroscopic dynamics, such as collective oscillations and synchronization.
%
The study of large-scale neural networks has either built on numerical simulations of full network models, consuming substantial computational costs or relied on heuristic models, exemplified by the Wilson-Cowan model \cite{WilsonCowan1972,WilsonCowan2021}. Heuristic models successfully reproduce the collective activity observed in physiological experiments, but disregard the contribution of individual neuron behaviour to this activity. 
%
Exact mean-field models have recently been developed using ideas from statistical physics, known as the Ott-Antonsen \cite{Ott2008} and Watanabe-Strogatz approaches \cite{WS1993}. These methods allow one to obtain low-dimensional dynamical systems for population-averaged variables that bridge the microscopic properties of individual neurons and the macroscopic dynamics of the neural network. Therefore, they are called next-generation neural mass models \cite{Byrne2020}. For a recent review of their development in neuroscience, please see \cite{Ashwin2016Rew} and \cite{Bick2020}.
%
In addition, time delays plaguing communication between neurons are an essential factor that cannot be ignored. Specifically, the time for a signal to travel from one neuron to the other through a chemical synapse can be up to 4 milliseconds \cite{Katz1965_SynapticDelay}. This is comparable to the characteristic time scales of the system in most situations and thus may have a significant influence on system dynamics. Such effects can be diverse. While it is well known that delays are associated with the emergence or prevention of oscillations \cite{Smith2011,reddy1998time}, delays can also control phase locking \cite{Coombes1997_DelayPhaseLock} and induce multi-stability \cite{ShayerCampbell2000_DelayMultistab, ChenCampbell2021_hysteresis}. 
To review the role of time delays in neural systems, we refer the reader to the papers \cite{campbell2007_delay,wu2011introduction,wu2022time,Ermentrout2009Delay}. 
%

In this paper, we focus on the impact of synaptic delays on the macroscopic dynamics of the neural network through exact mean-field reductions.
%
In particular, we follow the Ott-Antonsen theory \cite{Ott2008} and employ the Lorentzian ansatz \cite{Montbrio2015} to arrive at mean-field models for heterogeneous networks of Izhikevich neurons \cite{Izhikevich2003}, which take into account the mechanism of the spike frequency adaptation (SFA). 
For the network connections, we adopt a biologically realistic synaptic current model commonly used in neuroscience, but usually simplified outside this research field, e.g., \cite{Montbrio2015, Ratas2018_GammaDistr,Ferrara2023}. This simplification loses physiological accuracy to some extent. 
We also introduce a delay in the coupling to represent the delay in synaptic transmission. Fixed, homogeneous delay across the network is our main concern. These lead to a mean-field system composed of delayed differential equations.
Bifurcation analysis explore the impact on the macroscopic behaviours from important parameters, including  delay magnitude, adaptation intensity, and heterogeneity of the input current.
We conduct numerical continuation by employing standard software DDE-Biftool \cite{DDE-BIFTOOL} for the model with delay,  and XPPAUT \cite{XPPAUT} and MatCont \cite{Matcont} for the model with no delay, for comparison. 
To compare predictions of the mean-field model to the behaviour of the large network model, numerical simulations are carried out in Matlab (R2022b) \cite{MATLAB} and Julia (v1.8) \cite{Julia}.
%
Numerical bifurcation analysis allows us to deal with the more complex model that arises in our work due to the introduction of the realistic synaptic current expression and the SFA mechanism. In particular, it allows us to study new kinds of collective oscillations induced by higher-codimension bifurcations, such as population bursting induced by Torus (or Neimark–Sacker) bifurcation. 
%
The limit of heterogeneous applied current in the mean-field system was studied by \cite{PazoMontbrio2016, DevalleMontbrio2018_delay, Ratas2018_GammaDistr}, although this is in conflict with the Ott-Antonsen theory used to derive the neural mass model. To investigate this limit in our model, we perform perturbation analysis of the equilibrium points of the mean-field model and compare the bifurcation dynamics with the system having a dramatically weak heterogeneity.

This paper is organized as follows. Sec. 2 describes the Izhikevich neural network and the mean-field models derived from it. 
Sec. 3 includes the existence and the linear stability analysis of the equilibrium points for the mean-field system 
and the perturbation analysis when the heterogeneity of the input current is very weak. Sec. 4 is devoted to the numerical bifurcation analysis to investigate the impact of heterogeneity, adaptation intensity and synaptic delay on collective dynamics. Finally, Sec. 5 gives a summary and a brief discussion of the results. 

\section{Model description}

The Izhikevich neuronal model is an extension of the quadratic-integrate-and-fire (QIF) model to include a mechanism for spike frequency adaptation. 
It retains the central bifurcation properties of the QIF model, but exhibits a much larger class of bifurcations and associated solutions. Although not based on biophysical principles, experimental data can be used to fit the parameters so that the model accurately replicates neural spiking behaviour \cite{Izhikevich2007}.
The network of heterogeneous, all-to-all coupled  Izhikevich neurons is described by the following ordinary differential equations (ODEs),
\begin{equation}
\label{eq:network-Izh}
\begin{split}
v'_k(t) 
&=
v_k(v_k - \alpha)
-
w_k
+
I_{\mathrm{ext}} + \eta_k  
+
I_{\mathrm{syn},k} 
\\
w'_k(t)
&=
a(b v_k - w_k)
\\
\mathrm{if} \;\;
v_k 
& \geq
v_{\mathrm{peak}}, \;\;
\mathrm{then} \;
v_k \leftarrow v_{\mathrm{reset}}, 
w_k \leftarrow w_k + w_{\mathrm{jump}}
\end{split}
\end{equation}
for $k=1,\; 2,\; \dots,\; N$. Here, $'=d/dt$ denotes the time derivative, while $v_k$ represents the membrane potential of $k$th neuron and $w_k$ represents the adaptation current variable. The differential equations for $v_k$ and $w_k$ determine the neural activity leading to an action potential (voltage spike). When $v_k$ reaches the peak value $v_\mathrm{peak}$, a spike has occurred and $v_k$ is reset to $v_\mathrm{reset}$, while $w_k$ is incremented by an amount $w_\mathrm{jump}$, representing after-spike behaviour. This paper considers the limit $v_\mathrm{peak}=-v_\mathrm{reset}=\infty$. By introducing the change of variables $v=\tan(\theta/2)$, the model \eqref{eq:network-Izh} is transformed into an adaptive theta-neuron defined in the domain $\theta\in [-\pi, \pi)$ \cite{Izhikevich2007, ermentrout1986parabolic,Gutkin2022,Laing2018}.
Each neuron has three sources of input current: a common, possibly time-varying, component $I_\mathrm{ext}(t)$, a heterogeneous, quenched component $\eta_k$, drawn from a probability distribution, and a recurrent component representing input from other neurons via the synapses, $I_\mathrm{syn,k}$.

For the connection between neurons, we use the standard synaptic current model in neuroscience \cite{Ermentrout2010book}, 
\begin{equation}
\label{eq:synaptic_current}
I_{\mathrm{syn},k}(t)
=
g_\mathrm{syn}s_k(t)\big(e_r -v_k(t)\big)
=
g_\mathrm{syn}s(t)\big(e_r -v_k(t)\big)
\end{equation}
where $s_k$ is the synaptic gating variable, representing the proportion of ion channels open in the $k$th postsynaptic  neuron as the result of the firing in presynaptic neurons. The parameter $g_\mathrm{syn}$ is the maximum synaptic conductance which determines the strength of the coupling. The parameter $e_r$ is the reversal potential, which determines whether the synapses are excitatory or inhibitory, i.e, whether they increase or decrease the likelihood of firing of the postsynaptic neuron.
For an all-to-all coupling network, $s_k=s$ for all $k$ since every postsynaptic neuron receives the same summed input from all the presynaptic neurons \cite{Nicola2013bif}.
It is worth noting that some literature, e.g.,  \cite{Montbrio2015, Ratas2018_GammaDistr,Ferrara2023}, replace Eq.~\eqref{eq:synaptic_current} by  
$I_\mathrm{syn}= g s(t)$, where $g$ is constant. In this case, positive/ negative $I_\mathrm{syn}$ represents the excitatory/inhibitory synaptic coupling, respectively. 
%
%
However, this simplification ignores a physiological fact that the type of synaptic coupling is determined by the value of $e_r$ in Eq.~\eqref{eq:synaptic_current}, which does result in negative values of $I_\mathrm{syn}$ in the inhibitory network, but positive or negative $I_\mathrm{syn}$ in the excitatory network, depending on the sign of $(e_r-v_k)$.

Further, we use the single exponential model \cite{Ermentrout2010book} to govern the gating variable $s(t)$ with the equation,
\begin{equation}
\label{eq:single-exp-synapse_s_u}
s'(t) 
=
-s(t)/\tau_s
+
s_\mathrm{jump} u(t),
\end{equation}
where $\tau_s$ is the time constant, $s_\mathrm{jump}$ is the coupling strength and $u(t)$ represents the mean spike train injected from all presynaptic neurons, which is written as 
\begin{equation}
\label{eq:spike_train_nodelay}
\begin{split}
u(t)
&=
\frac{1}{N}
\sum_{l=1}^{N}\sum_{j \backslash t_l^j < t}\delta(t-t_l^j),
\\
r(t)
&=
\lim\limits_{N\to \infty}
u(t).
\end{split}
\end{equation}
Here, $\delta(t)$ is the Dirac delta function, $t_l^j$ is the time of the $j$th spike of the $l$th neuron, and when the number of neurons in the network goes to infinity, that is, $N \to \infty$, $u(t)$ is equivalent to the population firing rate $r(t)$, i.e., the population-averaged number of spikes per unit time \cite{Nicola2013bif}. 
It is easy to extend the single exponential synapse to other synaptic models, e.g., the double exponential synapse and the alpha synapse \cite{Ermentrout2010book, Nicola2013bif}. Note that since $s(t)$ is a proportion, it should be bounded by 1. This bound is implemented throughout the paper.

Next, we introduce the Lorentzian ansatz \cite{Montbrio2015}, which arises in the Ott-Antonsen theory \cite{Ott2008}. 
Considering that the current parameters $\eta_k$ obey a Lorentzian distribution of half-width $\Delta_\eta$, centered at $\bar \eta$,
\begin{equation}
\label{eq:Lorentzian_eta}
\mathcal{L}(\eta) 
=
\frac{1}{\pi}
\frac{\Delta_\eta}{(\eta - \bar \eta)^2 + \Delta_\eta^2},
\end{equation}
and following the specific approach in \cite{ChenCampbell2022_mf}, we can derive a system of ODEs in terms of the population's mean firing rate $r$, mean membrane potential $v$ and mean adaptation current $w$, given by
\begin{equation}
\label{eq:mf_izh_rvw}
\begin{split}
r'
& = 
\Delta_\eta/\pi 
+
2r v
-
\big(\alpha + g_{\mathrm{syn}}s \big)r
\\
v'
& =
v^2
- \alpha v
-
\pi^2 r^2
-
w
+
g_{\mathrm{syn}} s 
\big(
e_r - v
\big)
+
\bar \eta + I_{\mathrm{ext}}
\\
w'
&=
a
\left (
b 
v
-
w
\right )
+
w_{\mathrm{jump}} r
\end{split}
\end{equation}
Thus, together with Eqs. \eqref{eq:single-exp-synapse_s_u} and  \eqref{eq:spike_train_nodelay},  we arrive at a mean-field system to approximate the macroscopic behaviours of the Izhikevich neural network without synaptic delay in the thermodynamic limit, i.e., when $N\to \infty$.

If communication between neurons is not instantaneous, that is, there exists a delay between spike emission and reception, then for the $k$th neuron, the spike train resulting from other neurons obeys the following equation,
\begin{equation}
u_k(t) 
=
\frac{1}{N}
\sum_{l=1}^{N}\sum_{j \backslash \;t_l^j<t}
\delta(t-t_l^j - \tau_{l,k}),
\end{equation}
where $\tau_{l,k}$ is the synaptic delay for coupling between any $l$th and $k$th neurons.
For the simplicity, we assume the delay is homogeneous across the network and denote $\tau_{l,k}=D$ for all $l$ and $k$. Then, we have
\begin{equation}
\label{eq:spike_train_homo_delay}
u(t)
=
\frac{1}{N}
\sum_{l=1}^{N}\sum_{j \backslash \;t_l^j<t}
\delta(t-t_l^j - D).
\end{equation}
If the synaptic delay $\tau_{l,k}$ is heterogeneous and follows some probability density function $h(\tau)$ such that $h(\tau)d\tau$ represents the fraction of links with delays between $\tau$ and $\tau+d\tau$. In the thermodynamic limit, the mean spike train can be approximated by the integral in terms of $r(t)$ \cite{LeeOtt2009_delay},
\begin{equation}
\label{eq:spike_train_distr_delay}
u(t)
=
\int_0^\infty r(t-\tau)h(\tau)d\tau.
\end{equation}
In particular, $h(\tau) = \delta(\tau-D)$ corresponds to the homogeneous delay with $D$ on all links. Then, the mean spike train is transformed into 
\begin{equation}
\label{eq:spike_train_homo_delay_r}
u(t)=r(t-D).
\end{equation}
%

So far, we have developed the mean-field system of differential equations \eqref{eq:single-exp-synapse_s_u}, \eqref{eq:mf_izh_rvw} and \eqref{eq:spike_train_distr_delay} for the Izhikevich neural network with general distribution of synaptic delays. 
Note that all variables and parameters in this paper are dimensionless. To transform them into dimensional ones where parameters have physiological interpretation, one can refer to the paper \cite{ChenCampbell2022_mf}.


\section{Linear stability analysis}
\label{sec:pert_analysis}

In this section, we consider the general mean-field system, which is characterized by four macroscopic variables: $r(t)$, $v(t)$, $w(t)$ and $s(t)$.
The system has the same equilibrium point (EP) regardless of whether there is synaptic delay or not. 
We denote the EP as $(r_s, v_s, w_s, s_s)$. 

After some manipulations, we have
\begin{equation*}
\begin{split}
v_s
&=
\frac{J}{2} r_s
-
\frac{\Delta_\eta}{2\pi}\frac{1}{r_s}
+
\frac{\alpha}{2}
\\
w_s
&=
b v_s 
+
\frac{w_\mathrm{jump}}{a}r_s
\\
s_s
&=
\tau_s s_\mathrm{jump} r_s
\end{split}
\end{equation*}
where $J=g_\mathrm{syn}\tau_s s_\mathrm{jump}$
and $r_s$ satisfies the quartic equation 
\begin{equation}
\label{eq:mf_ep_r}
C_4 r_s^4 
+
C_3 r_s^3
+
C_2 r_s^2
+
C_1 r_s
+
C_0
=0.
\end{equation}
Here,
\begin{equation}
\label{eq:mf_ep_r_cof}
\begin{split}
C_4 
&=
J^2 + 4\pi^2,
\\
C_3 
&= 
2J 
(\alpha + b -2e_r)
+
4 w_\mathrm{jump}/a,
\\
C_2 
&= 
\alpha^2 + 2\alpha b - 4I_\mathrm{ext} -4\bar \eta,
\\
C_1 
&=
-2 b \Delta_\eta /\pi,
\\
C_0
&=
-\Delta_\eta^2/\pi^2.
\end{split}
\end{equation}
Note that for the biophysical interpretation, the population firing rate $r$ should be non-negative, and the proportion of open ion channels $s$ should be bounded by 1. So we have $0 \le r_s \le 1/(\tau_s s_\mathrm{jump})$.

Now consider the case of weak heterogeneity in the current parameter $\eta$, that is, $0 < \Delta_\eta = \epsilon << 1$ in Eq.~\eqref{eq:Lorentzian_eta}.  
The perturbation analysis in \ref{app:pert_EP} shows that the system has at most four biophysically relevant values for $r_s$, given by
\begin{equation}
\label{eq:mf_EP_r}
\begin{split}
r_s
&=
r_{s,0}^\pm
=
\frac{- C_3 
\pm
\sqrt{ C_3^2 - 4 C_4 C_2}}
{2 C_4}
\\
r_s
&=
r_{s,1}^\pm \cdot \epsilon
=
\frac{- \widetilde C_1
\pm
\sqrt{\widetilde C_1^2
-
4 \widetilde C_0 C_2}
}
{2C_2}
\cdot \epsilon
\end{split}
\end{equation}
where 
$\widetilde C_1=C_1/\epsilon=-2b/\pi$ and $\widetilde C_0= C_0/\epsilon^2=-1/\pi^2$.
The former two solutions can be expressed in parametric form in terms of $\bar \eta$ and $J$. 
The other two solutions are very close to zero. No solution with exactly zero firing rate exists, except when $\Delta_\eta = 0$ in Eq.~\eqref{eq:mf_izh_rvw}. 
However, we cannot take for granted that solutions found by setting $\Delta_\eta = 0$ in the mean-field model correspond to valid solutions for the full network. Since $\Delta_\eta = 0$ corresponds to the neurons in the network being strictly identical, the correct approach to derive the mean-field model is to 
resort to the Watanabe-Strogatz theory \cite{WS1993,WS1994}, rather than the Ott-Antonsen theory \cite{Ott2008} or the equivalent Lorentzian ansatz \cite{Montbrio2015} used in this paper.
However, our perturbation analysis of the mean-field model shows that the small EP $r_s \rightarrow 0$ as  $\Delta_\eta \rightarrow 0$. This implies that the case for the weak heterogeneity can be regarded as a smooth perturbation of the homogeneous one. This result provides a mathematical justification for some papers in the literature, e.g., \cite{Ratas2018_GammaDistr, Pazo2016, Devalle2018_delay}, where the analysis was built on $\Delta_\eta = 0$ in the Lorentzian-ansatz based mean-field model.

\begin{figure}
\centering
\includegraphics[width=\textwidth]{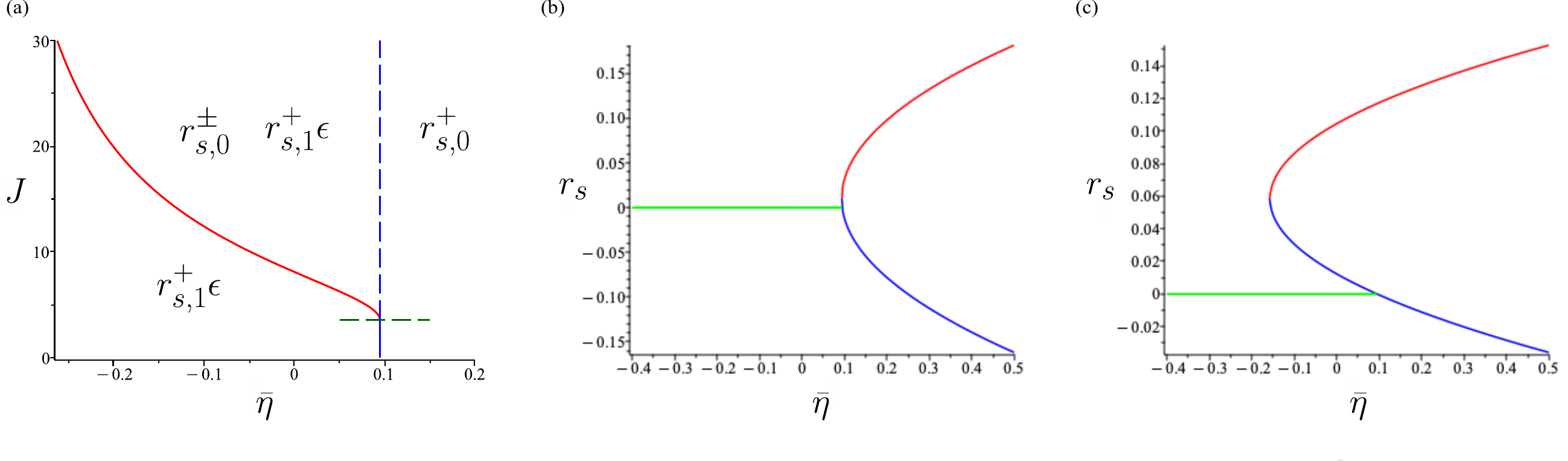}
\caption{Perturbation analysis of the mean-field system \eqref{eq:single-exp-synapse_s_u}, \eqref{eq:mf_izh_rvw} and \eqref{eq:spike_train_distr_delay} for the excitatory neural network with weak heterogeneity. 
(a): Existence of EPs with biophysically realistic value of $r_s$ in the $(\bar \eta, J)$ parameter space is determined by the curves defined from Eq.~\eqref{eq:mf_EP_r}.  Red corresponds to $C_3^2-4C_4 C_2=0$, blue to $C_2=0$, and green to $C_3=0$.
 Variation of $r_s$ with $\bar \eta$ when (b): $\Delta_\eta=10^{-4}$ and $J=3.94$,  
(c):  $\Delta_\eta=10^{-4}$ and $J=16.00$. Other parameter values are from Table \ref{tab:para-dimensionless}.}
\label{fig:EP_domain_E_pert}
\end{figure}

Further, from Eq.~\eqref{eq:mf_EP_r}, we can obtain regions in the parameter space where the different perturbation solutions for $r_s$ exist. A detailed analysis can be found in  \ref{app:pert_EP}. Here, we focus on an example, using parameter values from the literature.  Fig.~\ref{fig:EP_domain_E_pert} (a) illustrates regions of existence in the $(\bar \eta, J)$ plane. 
For the excitatory neural network where the dimensionless value of the reversal potential $e_r=1$, the system has only one EP with $r_s=r_{s,0}^+>0$ on the right of the blue line, three EPs with $r_s = r_{s,0}^{\pm}>0$ and $r_s = r_{s,1}^+\epsilon \approx 0$ in the region bounded by the red and blue curves, and only one EP with $r_s=r_{s,1}^+\epsilon \approx 0$ in the left region.
With $b=-0.0062$, the region of existence for the EP with $r_s = r_{s,1}^- \epsilon$ is extremely narrow. Thus it is not shown. 
Fig.~\ref{fig:EP_domain_E_pert} (b) and (c) show the variation of $r_s$ with respect to $\bar \eta$ for different values of $J$. 
In (b) $J=g_\mathrm{syn}\tau_ss_\mathrm{jump}=3.94$, i.e., $g_\mathrm{syn}=1.2308$ as used in \cite{ChenCampbell2022_mf}. Only the EP with $r_s=r_{s,1}^+\epsilon \approx 0$ (green) exists for $\bar \eta < 0.0946$. Two EPs, with $r_s=r_{s,0}^{\pm}>0$, exist in an extremely narrow region with $\bar \eta > 0.0946$ (red and tiny part of blue curve). For larger $\bar\eta$ only the EP with $r_s=r_{s,0}^+>0$ (red) exists.  
In (c) $J=16.00$, i.e., $g_\mathrm{syn}=5$. For $\bar \eta > 0.0946$ only the EP with $r_s=r_{s,0}^+>0$ (red) exists. For $\bar \eta <-0.1570$ only the EP with $r_s=r_{s,1}^+\epsilon \approx 0$ (green) exists.  Between the two ranges, there are three EPs.
For the inhibitory neural network where $e_r<0$, only one EP with $r_s=r_{s,0}^+>0$ exists on the right of the blue line in Fig.~\ref{fig:EP_domain_E_pert} (a), that is, $\bar \eta > 0.0946$ and only one EP with $r_s=r_{s,1}^+\epsilon \approx 0$ on the left.
Fig.~\ref{fig:EP_domain_E_pert} was created using Maple \cite{maple}. The existence of the EPs was numerically verified by XPPAUT \cite{XPPAUT} (not shown here).


Next, we study the linear stability of the EPs. Define small deviations $\delta r = r - r_s$,
$\delta v = v-v_s$,
$\delta w = w - w_s$ and $\delta s = s - s_s$, and linearize the mean-field equations around the EP. By looking for the solution of the linearized equations in the form $(\delta r, \delta v, \delta w, \delta s) \propto \exp(\lambda t)$, we obtain the characteristic equation
\begin{equation}
\label{eq:char_eq_mf_delay}
\Lambda(\lambda)
=
P(\lambda)H^{-1}(\lambda)
+
Q(\lambda)
\end{equation}
where $H(\lambda)$ is the Laplace transform of $h(\tau)$, that is,
\begin{equation*}
H(\lambda)
=
\int_0^\infty
e^{-\lambda \tau}
h(\tau)d\tau
\end{equation*} 
and
\begin{equation*}
\begin{split}
P(\lambda)
&=
\Big [
(\lambda + a)(\lambda + K)^2 
+
ab (\lambda + K)
+ 
4\pi^2 r_s^2 (\lambda + a)
+
2  w_\mathrm{jump} r_s
\Big ]
(\tau_s\lambda + 1) 
\\
Q(\lambda)
&=
\Big [
J r_s 
(\lambda + K)(\lambda + a)
-
2 J r_s   
(e_r - v_s)
(\lambda + a)
+ 
ab J r_s  
\Big ]
\end{split}
\end{equation*}
Here, $J = \tau_s g_\mathrm{syn} s_\mathrm{jump}$ and $K = \alpha + J r_s - 2 v_s$. See \ref{app:char_EP_HP} for more details.
We have the following general information about the stability region.
\begin{itemize}
\item $H(0) = 1$
\item for real roots of $\Lambda(\lambda) = 0$, if $\lambda \ge 0$, then $e^{-\lambda \tau} \le 1$, and thus $H(\lambda) \le 1$.
\item for some parameter values, the characteristic equation can have a zero root for any distribution, that is, $\Lambda(0) = P(0)\cdot 1 + Q(0) = 0$.
\end{itemize}
In addition, we can derive the potential Andronov-Hopf bifurcation curves in a parametric form by imposing the condition of marginal stability $\lambda = i\omega$ on the characteristic equation \eqref{eq:char_eq_mf_delay}. 
Denote $H^{-1}(i\omega) = A(\omega) + i B(\omega)$.  Separating the real and imaginary parts of the characteristic equation yields
\begin{equation}
\label{eq:char_eq_hopf_mf_delay}
\begin{split}
(R_2 r_s^2 + R_1 r_s + R_0)A
-
(D_2 r_s^2 + D_1 r_s + D_0)B\omega
+
(\hat R_2 r_s^2 + \hat R_1 r_s)
&= 0
\\
(R_2 r_s^2 + R_1 r_s + R_0)B
+
(D_2 r_s^2 + D_1 r_s + D_0)A\omega
+
(\hat D_2 r_s^2 + \hat D_1 r_s)\omega
&= 0
\end{split}
\end{equation}
where $R_j$, $\hat R_j$, $D_j$ and $\hat D_j$ ($j=0, 1, 2$) are parametric coefficients with the expressions shown in \ref{app:char_EP_HP}. The appendix also includes the expressions of $H^{-1}(i\omega)$ with and without synaptic delay. 
We determine $r_s$ as a function of parameters from Eq.~\ref{eq:mf_EP_r} and substitute it into Eq.~\eqref{eq:char_eq_hopf_mf_delay}. Then we can derive the value of one parameter and $\omega$ at which a Hopf bifurcation may occur. Alternatively, we could solve for two parameters as parametric equations in $\omega$ to get two-parameter 
curves. To fully verify the existence of Hopf bifurcation requires checking the transversality and nondegeneracy conditions~\cite{Smith2011}. Given the complex dependence of the EPs, Eq.~\eqref{eq:mf_ep_r_cof}, and hence of the coefficients of the characteristic equation on the model parameters, these conditions are difficult to check for the general model. Hence we turn to numerical methods for the rest of the paper.


\section{Numerical bifurcation analysis}

In this section, we focus on the homogeneous delay scenario, that is, the delays in the synaptic transmission between any pair of neurons are the same. 
We perform numerical bifurcation analysis on the mean-field model and mainly investigate the impact of three factors on the 
macroscopic dynamics: the heterogeneity of the current $\Delta_\eta$, the level of adaptation $w_\mathrm{jump}$ and the synaptic delay $D$.
 Our results are obtained using standard continuation packages: XPPAUT \cite{XPPAUT} and MatCont \cite{Matcont} for the system with zero delay, and DDE-Biftool \cite{DDE-BIFTOOL} when the delay is nonzero. To verify and extend these results, numerical simulations are carried out in Matlab (R2022b) \cite{MATLAB} and Julia (v1.8) \cite{Julia}. The results reported here complement those based on the linear stability analysis discussed in the previous section.

\begin{table}[ht!]
\centering
\renewcommand{\arraystretch}{1.2} 
\begin{tabular}{cccc}
\hline
\hline
Parameter & Value & Parameter & Value\\
\hline
  $\alpha$
  & $0.6215$
  & $\tau_s$
  & $2.6$ \\
  $a$
  & $0.0077$ 
  & $b$
  & $-0.0062$ \\
  $I_\mathrm{ext}$
  & 0
  & $s_{\mathrm{jump}}$
  & $1.2308$  \\
  $g_\mathrm{syn}$
  & $1.2308$ 
  & $w_{\mathrm{jump}}$ 
  & $0.0189$  \\
  $e_{r,E}$ 
  & 1
  & $e_{r,I}$ 
  & -0.1538 \\
\hline
\hline
\end{tabular}
\caption{Dimensionless parameter values for the Izhikevich neural network ($e_{r,E}/e_{r,I}$ for excitatory/inhibitory network)}
\label{tab:para-dimensionless}
\end{table}

\begin{table}[ht!]
\centering
\renewcommand{\arraystretch}{1.2} 
\begin{tabular}{ccc}
\hline
\hline
Parameter & Dimensionless value & Dimensional Value\\
\hline
  $\bar \eta$
  & $0-0.5$
  & $0-5000$ pA \\
  $\Delta_\eta$
  & $0-0.1$ 
  & $0-1056$ pA \\
  $g_{\mathrm{syn}}$
  & $0-5$
  & $0-800$ nS \\
  $w_\mathrm{jump}$
  & $0-0.1$
  & $0-1056$ pA\\
  $D$
  & $0-20$ 
  & $0-30$ ms \\
\hline
\hline
\end{tabular}
\caption{Physiologically plausible ranges of parameters for bifurcation analysis}
\label{tab:para-bif}
\end{table}

Table \ref{tab:para-dimensionless} gives dimensionless values of the model parameters used in the paper unless otherwise indicated in a figure caption. The corresponding dimensional values and scaling relations are given in \cite{ChenCampbell2022_mf}.
All values are taken from \cite{Nicola2013bif} for the excitatory coupled neural network (E-net), which were initially fit by \cite{Dur-e-Ahmad2012} to hippocampal CA3 pyramidal neuron data from \cite{Hemond2008}. The parameters for the inhibitory network (I-net) are the same except that the reversal potential ($E_r=-75$ mV, that is, $e_r=-0.1538$) is taken from \cite{Rich2020}. 
To perform the bifurcation analysis, we consider the physiologically plausible ranges of parameter values \cite{Nicola2013hetero,Buszaki2006,Ermentrout2009Delay}, as shown in Table \ref{tab:para-bif}.

\subsection{Typical macroscopic behaviours}

\begin{figure}[ht!]
\centering
\includegraphics[width=\textwidth]{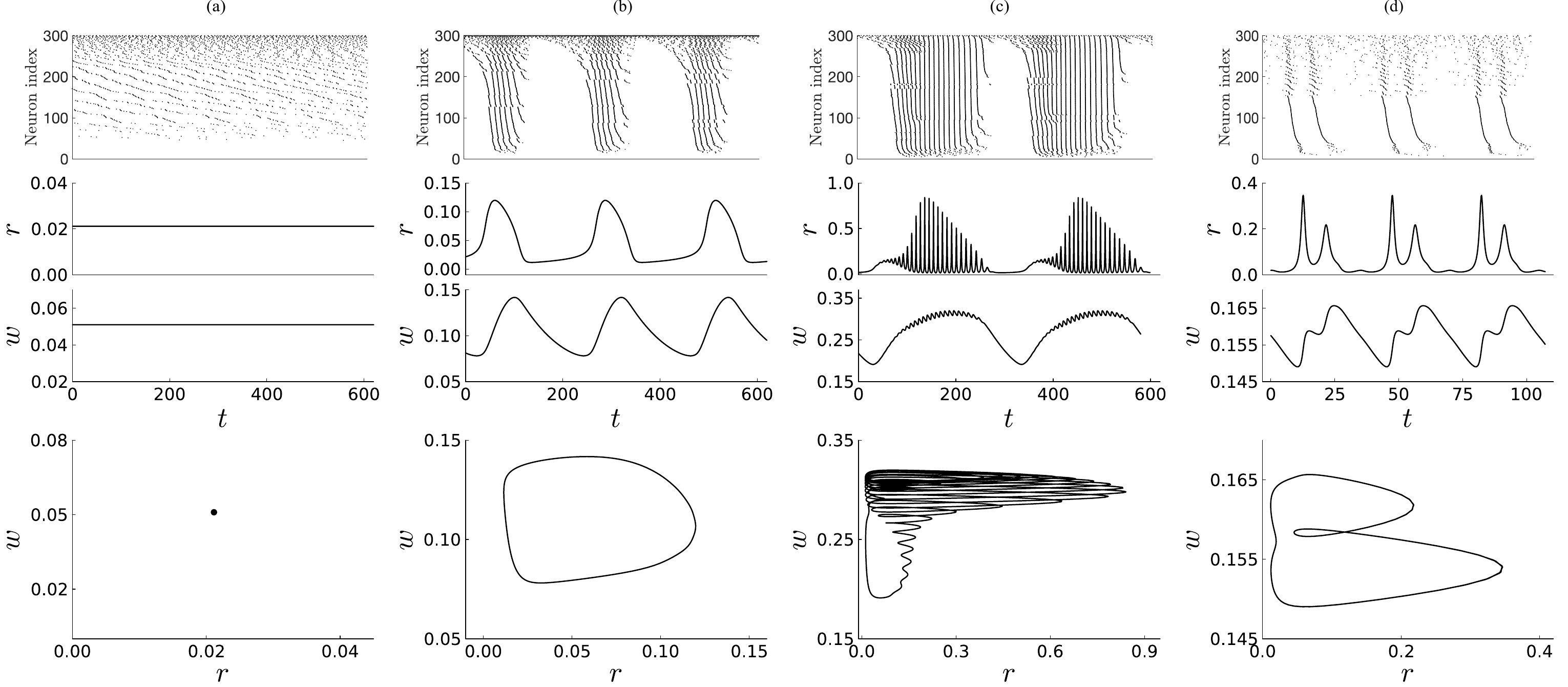}
\caption{Examples of macroscopic behaviours. The first row shows raster plots of 300 neurons randomly selected from 5000 neurons of the network \eqref{eq:network-Izh}, \eqref{eq:synaptic_current}, \eqref{eq:single-exp-synapse_s_u} and \eqref{eq:spike_train_homo_delay}. Dots correspond to firing events and neurons are arranged in order of increasing current $\eta_k$. The last two rows give time evolution and phase portraits of $r(t)$ and $w(t)$ of the delayed mean-field equations \eqref{eq:single-exp-synapse_s_u}, \eqref{eq:mf_izh_rvw} and \eqref{eq:spike_train_homo_delay_r}.
Parameter values: Column (a) (E-net): $D=1$, $g_\mathrm{syn}=0.2$, $w_\mathrm{jump}=0.0189$, $\bar \eta=0.12$ and $\Delta_\eta=0.02$;
Column (b) (E-net): $D=1$, $g_\mathrm{syn}=1$, $w_\mathrm{jump}=0.0189$, $\bar \eta=0.12$ and $\Delta_\eta=0.02$;
Column (c) (E-net): $D=4$, $g_\mathrm{syn}=1.1$, $w_\mathrm{jump}=0.025$, $\bar \eta=0.25$ and $\Delta_\eta=0.02$;
Column (d) (I-net): $D=14$, $g_\mathrm{syn}=1$, $w_\mathrm{jump}=0.0189$, $\bar \eta=0.4$ and $\Delta_\eta=0.02$. 
}
\label{fig:TimeSeries_RasterPlot}
\end{figure}

Before proceeding to the bifurcation analysis, let us consider some typical time evolutions of the models, shown in 
Fig.~\ref{fig:TimeSeries_RasterPlot}. The raster plots are obtained from numerical simulation of the full neural network model~\eqref{eq:network-Izh}, and while the time series of the macroscopic variables are from integrating the corresponding mean-field equations \eqref{eq:mf_izh_rvw}.
%
%
We start with the column Fig.~\ref{fig:TimeSeries_RasterPlot} (a). The raster plot shows that a small group of neurons with low current remain quiescent, while the rest exhibit asynchronous tonic firing with different frequencies determined by the current values. From the macroscopic viewpoint, the system finally settles on an EP with fixed $r(t)$, $w(t)$, corresponding to a point in the $(r,w)$ phase plane. 
The column Fig.~\ref{fig:TimeSeries_RasterPlot} (b) shows the dynamics where regular collective oscillations (COs) are observable. In the network, neurons undergo bursting characterized by alternation between silent and active states. During a burst, neurons spike asynchronously. Neurons with the higher current $\eta$ fire followed by those with lower one. The mean-field system converges to a periodic orbit (PO) with a unique oscillation period. 
In Fig.~\ref{fig:TimeSeries_RasterPlot} (c), neurons also exhibit bursting behaviour. The difference is that during a burst, most neurons fire synchronously, manifested by straight lines in the raster plot, instead of the    ``travelling waves" seen in Fig.~\ref{fig:TimeSeries_RasterPlot} (b). The mean-field system exhibits quasiperiodic behaviour 
characterized by fast oscillations with a slowly varying envelope. 
These COs involve two oscillation periods, the slow one responsible for the population bursting rhythm and the fast corresponding to the spiking period of the individual neurons. 
This kind of CO has been called slow-fast nested oscillations or population bursting from the macroscopic perspective in analogy with the similar activity of single neurons.
Another interesting CO is shown in  Fig.~\ref{fig:TimeSeries_RasterPlot} (d). In this case, a subpopulation of neurons burst, alternating with short and long quiescent periods. The limit cycle in the mean variables of $r(t)$ and $w(t)$, as seen in the time series and phase portrait, is similar to those that arise in period-doubling bifurcations.

In summary, we present four dynamical regimes occurring in our mean-field system.
The first two dynamics are common phenomena observed in the neural network, no matter whether we consider the delay in spike transmission or not.
The last two regimes are new and more complex COs found in our model by incorporating the mechanism of SFA and synaptic delay in a single population of neurons. 
In order to better understand these macroscopic behaviours and determine how they transition between each other, we turn to numerical bifurcation analysis in the following sections.

\subsection{Bifurcation of neural network without synaptic delay}

\begin{figure}[ht!]
\centering
\includegraphics[width=\textwidth]{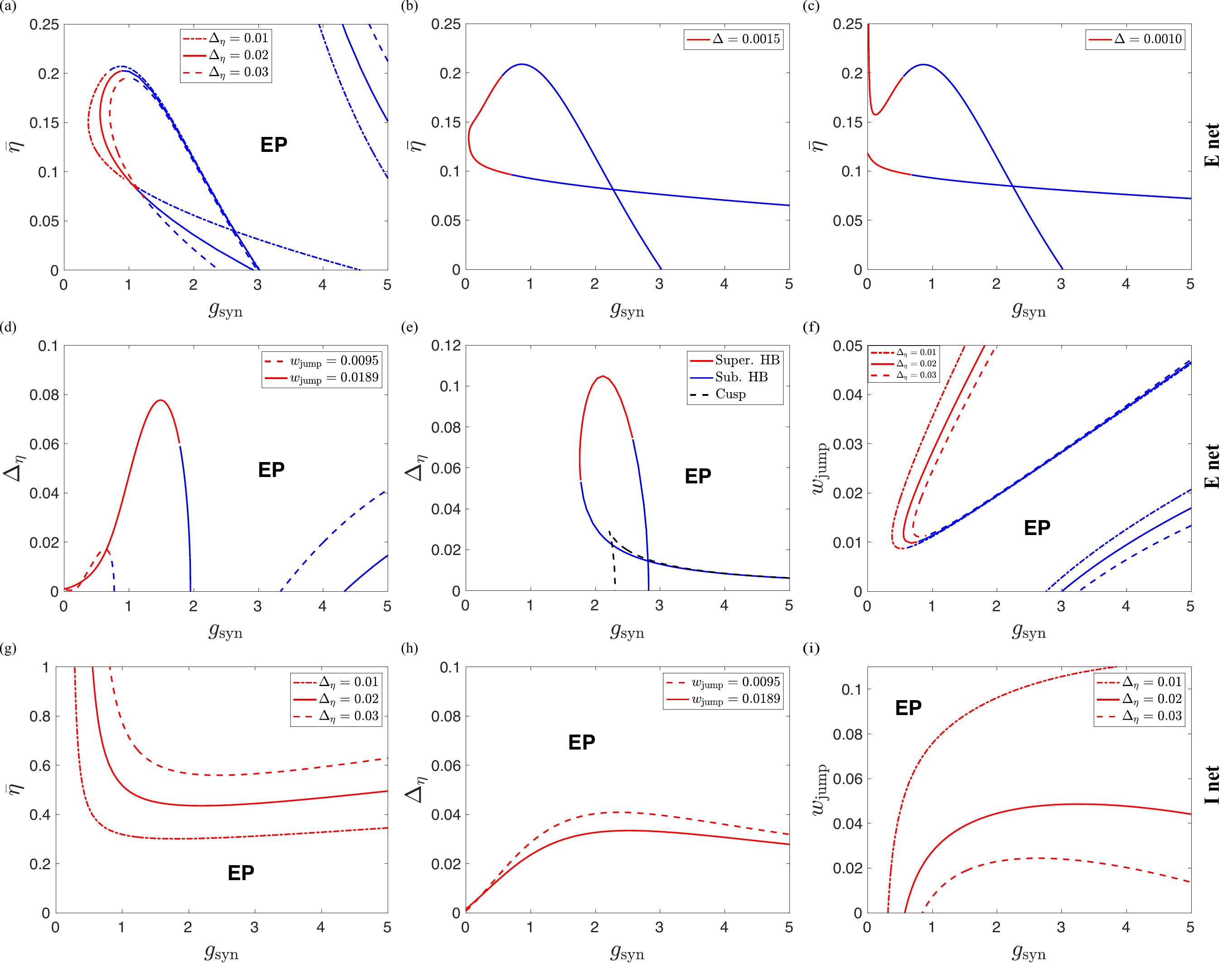}
\caption{Effect of heterogeneity and adaptation on the excitatory (first two rows) and inhibitory (last row) neural networks without synaptic delay. 
Red/blue: supercritical/subcritical Hopf boundaries.
Parameter values: (a)-(c)  $w_\mathrm{jump}=0.0189$, (d) $\bar \eta=0.12$, (e) $\bar \eta=0.02$, (f) $\bar \eta=0.12$, (g) $w_\mathrm{jump}=0.0189$, (h)-(i) $\bar \eta=0.6$. 
}
\label{fig:Effect_hw_wjump_NoDelay}
\end{figure}

Looking for Andronov-Hopf bifurcations is the first step of bifurcation analysis to separate collective oscillations from stationary states (i.e., EPs).
For comparison, we begin with the neural network without synaptic delay Eqs.~\eqref{eq:network-Izh}-\eqref{eq:spike_train_nodelay}. The corresponding mean-field system is composed of the ordinary differential equations \eqref{eq:single-exp-synapse_s_u}, \eqref{eq:spike_train_nodelay} and \eqref{eq:mf_izh_rvw}.
Hopf boundaries are shown in Fig.~\ref{fig:Effect_hw_wjump_NoDelay} in different pairs of parameter planes.

First, we explore the effect of the heterogeneity of the current $\Delta_\eta$.
During the derivation of mean-field approximations, we assume that the current $\eta$ follows a Lorentzian distribution \eqref{eq:Lorentzian_eta} with a centre at $\bar \eta$ and a half-width $\Delta_\eta$. The value of $\Delta_\eta$ determines the level of heterogeneity of the current injected into each neuron. $\Delta_\eta=0$ implies the homogeneous network, where each neuron is given the same current $\eta$. The larger $\Delta_\eta$ is, the more heterogeneous the network is.  

For the excitatory neural network 
we start with Hopf bifurcation in the $(g_\mathrm{syn},\bar \eta)$ plane for different values of $\Delta_\eta$, shown in Fig.~\ref{fig:Effect_hw_wjump_NoDelay} (a).
Two Hopf branches appear. One looks like a straight line in the biological range of parameter values, with the Hopf being subcritical (blue); the other is a closed loop and exhibits two codimension-2 Bautin (or generalized Hopf) bifurcations, with the Hopf being supercritical (red) between these points and subcritical (blue) outside.
In the region bounded by two Hopf branches, the mean-field system settles on an EP, where neurons exhibit asynchronous tonic firing, similar to the behaviour shown in Fig.~\ref{fig:TimeSeries_RasterPlot} (a). 
Inside the closed loop and in the top right corner, a CO emerges and neurons exhibit bursting behaviours. The behaviour is similar to that shown in  Fig.~\ref{fig:TimeSeries_RasterPlot} (b), although the oscillation frequency depends on the particular parameter values where the Hopf bifurcation occurs.
Additionally, note that the closed curves with different values of $\Delta_\eta$ appear to intersect at a point $(g_{syn,c},\bar\eta_c)\approx(1,0.09)$. For $\bar \eta > \bar\eta_c$, the $g_\mathrm{syn}$ value of the Hopf closed loop associated with the left boundary moves to the right when $\Delta_\eta$ increases, while the right boundary does not change appreciably. This leads to the CO area getting smaller as $\Delta_\eta$ grows. 
The same phenomenon occurs at the top right corner.
Consistent results are found in Fig.~\ref{fig:Effect_hw_wjump_NoDelay} (d) where $\bar \eta = 0.12>0.09$.
On the contrary, the CO region gets bigger as $\Delta_\eta$ grows in the region where $\bar \eta<\bar\eta_c$.
Consistent results are shown in Fig.~\ref{fig:Effect_hw_wjump_NoDelay} (e) for the range $\Delta_\eta<0.06$ approximately. However, the opposite effect appears when $\Delta_\eta>0.06$. Here, $\bar \eta = 0.02<\bar\eta_c$. 
Therefore, heterogeneity of the current $\eta$ either makes neurons' bursting behaviour more robust (bigger CO area) or less (smaller CO area), depending on the different region of parameter space the system lies in.
Our findings may help explain the contradictory conclusions obtained in \cite{Vladimirski2008} and \cite{Nicola2013hetero}, where the former claimed that heterogeneity favours the emergence of bursting as opposed to hinders, as the latter found. Although it is hard to directly compare our model with that in \cite{Vladimirski2008}, we can compare with the paper \cite{Nicola2013hetero} since both of us start from the same neural network model and use equivalent parameter values. 
Because of the different mean-field modelling strategy, the paper \cite{Nicola2013hetero} only considered the region where the current is larger than $1000$ pA (equivalent dimensionless value $\bar \eta > 0.0947$).  Our results in this range are consistent:  the emergence of COs benefits from higher heterogeneity of the current.

In addition, we note that 
in Fig.~\ref{fig:Effect_hw_wjump_NoDelay} (a) for the case when $\Delta_\eta = 0.01$, the left Hopf branch crosses the right one, forming another region where the system is attracted to the EP as well. This is supported by Fig.~\ref{fig:Effect_hw_wjump_NoDelay} (e), where the system undergoes cusp bifurcations, implying the presence of a hysteresis phenomenon which is often accompanied by bistability \cite{ChenCampbell2021_hysteresis}.
Finally, we consider the effect of weak heterogeneity. As $\Delta_\eta$ decreases, see Fig.~\ref{fig:Effect_hw_wjump_NoDelay} (b), the two-parameter Hopf curves move to the left, but the CO region remains qualitatively similar to Fig.~\ref{fig:Effect_hw_wjump_NoDelay} (a) for $\Delta_\eta$ large enough. 
For $\Delta_\eta$ close enough to zero, however, the closed Hopf branch splits with the lower part limiting on the $\bar\eta$ axis and the upper part approaching large current values as $g_\mathrm{syn}$ approaches zero. This scenario represents a population of neurons with extremely weak or zero coupling ($g_\mathrm{syn}<0.2$) and nearly homogeneous injected currents ($\Delta_\eta \approx 0$). 
%

For the inhibitory neural network, as shown in the last row of Fig.~\ref{fig:Effect_hw_wjump_NoDelay}, the bifurcation diagram is relatively simpler. Only supercritical Hopf bifurcations (red) are observable for the present choice of parameter values. 
COs emerge for sufficiently large  $\bar \eta$, (see Fig.~\ref{fig:Effect_hw_wjump_NoDelay} (g)) and almost any value of $g_\mathrm{syn}$ (see Fig.~\ref{fig:Effect_hw_wjump_NoDelay} (h) and (i) ). Increasing the heterogeneity in the applied current via $\Delta_\eta$ does not favour neurons' bursting behaviour, as shown in Fig.~\ref{fig:Effect_hw_wjump_NoDelay} (g)-(i). A higher $\bar \eta$ or a lower $w_\mathrm{jump}$ is required to overcome the increasing $\Delta_\eta$ to generate COs (see Fig.~\ref{fig:Effect_hw_wjump_NoDelay} (g) and (i)).

Next, we investigate the effect of the adaptation intensity parameter $w_\mathrm{jump}$.
The relevance of adaptation for the emergent collective dynamics can be appreciated by considering the bifurcation diagram in the $(g_\mathrm{syn}, w_\mathrm{jump})$ plane, as shown in  Fig.~\ref{fig:Effect_hw_wjump_NoDelay} (f) for the excitatory neural network and (i) for the inhibitory case.
In the excitatory case, COs emerge for sufficiently large $w_\mathrm{jump}$ or $g_\mathrm{syn}$. A generalized Hopf point occurs, separating supercritical (red) from subcritical (blue) Hopf bifurcations. The adaptation favours the emergence of COs with the CO region expanding as $w_\mathrm{jump}$ increases.
For the inhibitory case, COs appear even at $w_\mathrm{jump}=0$ with medium values of $g_\mathrm{syn}$. Adaptation in the inhibitory network has the opposite effect to the excitatory case. The interval of $g_\mathrm{syn}$ where COs occur shrinks as $w_\mathrm{jump}$ grows, thus adaptation hinders the formation of COs.

To understand the effects of the current heterogeneity parameter $\Delta_\eta$ and the adaptation parameter $w_\mathrm{jump}$, we consider the principle underlying spiking in the dynamic equation of the membrane potential \eqref{eq:network-Izh}. 
The emergence of COs in the synaptically coupled neural network is due to a balance between the intrinsic ($\eta$) and external ($I_\mathrm{ext}$) currents, which cause neurons to spike, the slow adaptation current $w$, which terminates spiking, and the synaptic current ($I_\mathrm{syn}$), which favours spiking in the excitatory network because it is positive most of the time, but hinders spiking in the inhibitory network as it is always negative. 
%
%
Larger heterogeneity implies the excitatory drive $\eta$ is distributed over a wider range of values. If the mean of the current $\bar \eta$ is sufficiently large, e.g., $\bar \eta=0.12$ in Fig.~\ref{fig:Effect_hw_wjump_NoDelay} (d), larger $\Delta_\eta$ indicates more neurons have  input currents too small to invoke spiking, resulting in a smaller bursting area. 
If $\bar \eta$ is relatively small, e.g., $\bar \eta=0.02$ and $\Delta_\eta\in[0.02, 0.06]$ in Fig.~\ref{fig:Effect_hw_wjump_NoDelay} (e), bigger $\Delta_\eta$ indicates more neurons have large enough currents to spike, resulting in a bigger bursting area. 
%
The synaptic current $I_\mathrm{syn}$ has different effects in the excitatory and inhibitory networks. As shown in Eq.~\ref{eq:synaptic_current}, the sign of $I_\mathrm{syn}$ is determined by the value of the membrane potential $v(t)$ relative to the reversal potential $e_r$. Thus it can be positive or negative in the excitatory network but is always negative in the inhibitory case.  Consequently, the bifurcation diagrams in the excitatory case are more complicated than those in the inhibitory one.
In the excitatory network, the primary source of inhibition arises from adaptation. The interaction between excitatory and inhibitory sources leads to multiple areas for COs, e.g., Fig.~\ref{fig:Effect_hw_wjump_NoDelay} (a) and (d), including (f), where even at $w_\mathrm{jump}=0$, strong coupling (approximately $g_\mathrm{syn}>3$) introduces an inhibition effect.
In the inhibitory case, the negative synaptic current and the adaptation both act to suppress COs. Thus, COs appear even at $w_\mathrm{jump}=0$ and disappear for sufficiently large $w_\mathrm{jump}$, e.g., Fig.~\ref{fig:Effect_hw_wjump_NoDelay} (h) and (i).

\subsection{Bifurcation of neural network with synaptic delay}

\begin{figure}[ht!]
\centering
\includegraphics[width=\textwidth]{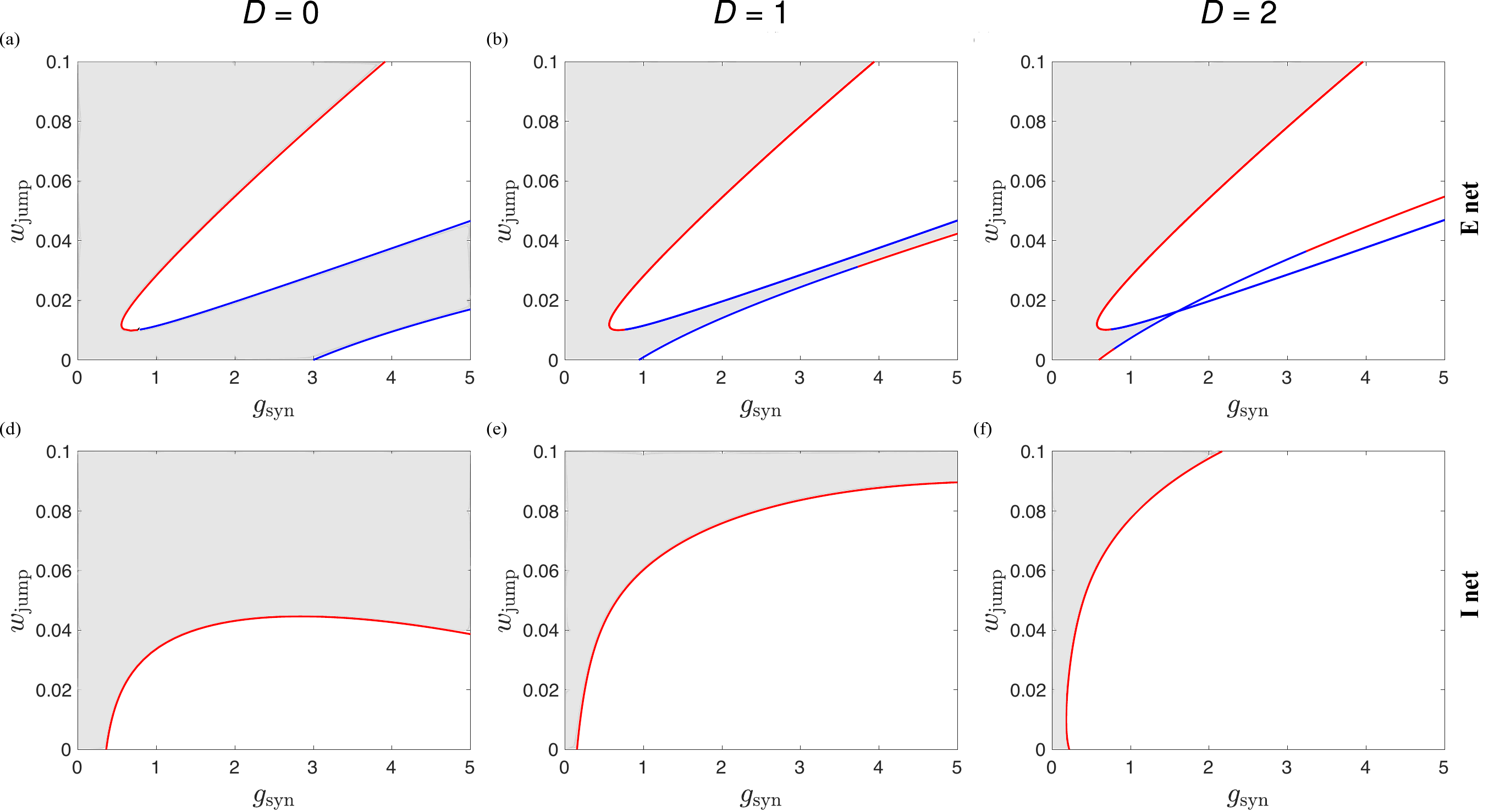}
\caption{Effect of delay on excitatory (1st row) and inhibitory (2nd row) neural networks. Boundaries of Hopf bifurcation (red/blue for supercritical/subcritical) are shown in the $(g_\mathrm{syn}, w_\mathrm{jump})$ plane for different values of delay $D=0$ (1st column), $D=1$ (2nd column) and $D=2$ (3rd column). Grey region: stable EP.
Parameter values: $\bar \eta=0.12$, $\Delta_\eta=0.02$ for the excitatory network, $\bar \eta=0.4$, $\Delta_\eta=0.01$ for the inhibitory network.
}
\label{fig:Effect_Delay_Discrete}
\end{figure}

In this section, we consider the network with constant, homogeneous synaptic delay. The corresponding network model is composed of \eqref{eq:network-Izh}-\eqref{eq:single-exp-synapse_s_u} and \eqref{eq:spike_train_homo_delay} and its mean-field system consists of delayed differential equations \eqref{eq:single-exp-synapse_s_u}, \eqref{eq:mf_izh_rvw} and \eqref{eq:spike_train_homo_delay_r}.
We start with an investigation of Hopf bifurcations. Fig.~\ref{fig:Effect_Delay_Discrete} shows Hopf  boundaries reported in the $(g_\mathrm{syn}, w_\mathrm{jump})$ plane for different values of delay $D$.  
Generally, delay favours the emergence of COs and generates new dynamics.
Specifically, in the inhibitory network (2nd row of Fig.~\ref{fig:Effect_Delay_Discrete}), the CO areas (white) under the supercritical Hopf curves (red) expand with the increase of $D$.
In the excitatory case (1st row of Fig.~\ref{fig:Effect_Delay_Discrete}), the delay not only promotes COs by the expansion of the white regions, but also enriches the collective dynamics.
In particular, as the delay increases, the Hopf branch at the lower right corner moves towards the upper one, shrinking the EP area (grey), creating a Hopf-Hopf bifurcation. A Bautin bifurcation also moves into the parameter range considered. 
Numerical simulations of the macroscopic variables show that the mean-field system with $D=0$ exhibits slow oscillations in the upper white region and fast oscillations in the lower white region, similar to the behaviour shown in  Fig.~\ref{fig:TimeSeries_RasterPlot} (b), but with different periods.
When $D$ increases to 2, these two dynamics persist in the upper and lower white regions. However, in the middle white region created by the intersection of the two Hopf curves the system can exhibit slow-fast nested COs, similar to those shown in  Fig.~\ref{fig:TimeSeries_RasterPlot} (c).

\begin{figure}
\centering
\includegraphics[width=\textwidth]{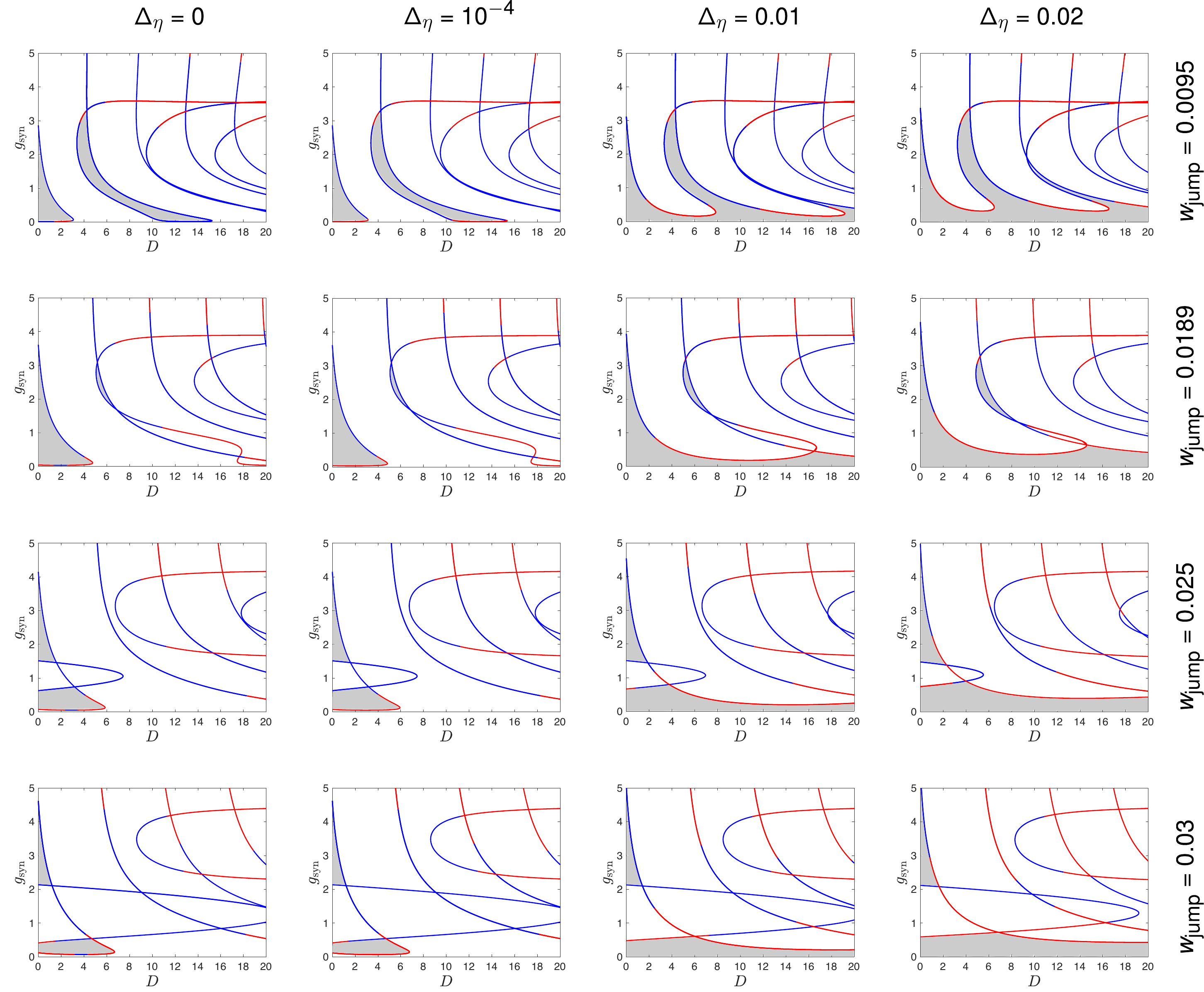}
\caption{Excitatory neural network with homogeneous synaptic delay. Hopf curves (red/blue for supercritical/subcritical)  are shown in the $(D, g_\mathrm{syn})$ plane for various values of $\Delta_\eta$ and $w_\mathrm{jump}$ when $\bar \eta=0.25$. Grey region: stable EP.}
\label{fig:Effect_Delay_gsyn_hw_wjump_filled}
\end{figure}

\begin{figure}
\centering
\includegraphics[width=\textwidth]{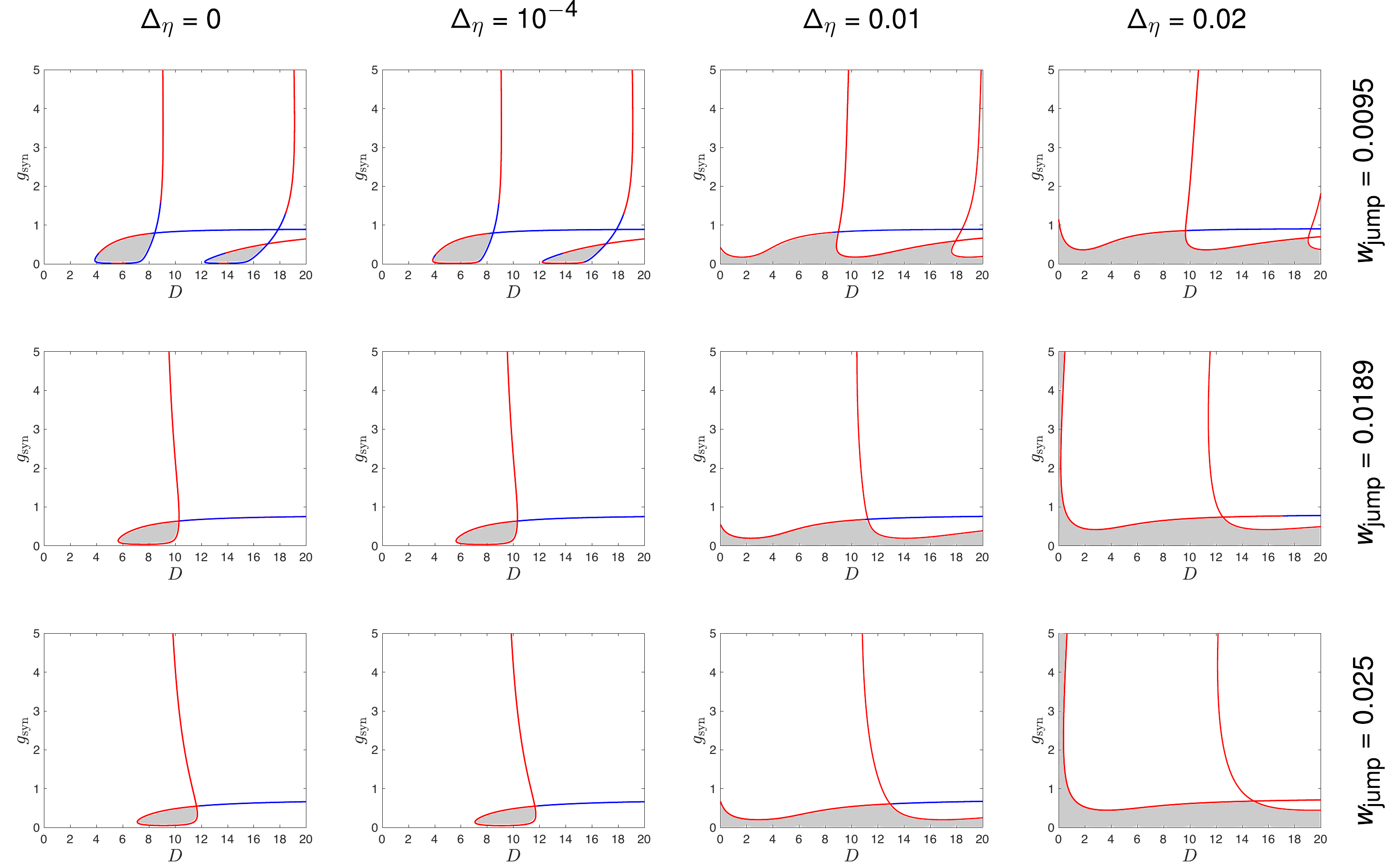}
\caption{Inhibitory neural network with homogeneous synaptic delay. Hopf curves are shown in the $(D, g_\mathrm{syn})$ plane for different values of $\Delta_\eta$ and $w_\mathrm{jump}$ when $\bar \eta=0.4$. Grey region: stable EP.}
\label{fig:Effect_Delay_gsyn_hw_wjump_I_net_filled}
\end{figure}

To further study the impact of the input current heterogeneity and adaptation on the time-delayed mean-field system, we plot the Hopf bifurcation curves in the $(D,g_\mathrm{syn})$ plane for different values of $\Delta_\eta$ and $w_\mathrm{jump}$, as shown in Fig.~\ref{fig:Effect_Delay_gsyn_hw_wjump_filled} for the excitatory network and Fig.~\ref{fig:Effect_Delay_gsyn_hw_wjump_I_net_filled} for the inhibitory case.
The adaptation strength increases from top to bottom with $w_\mathrm{jump}=0.0095$ (100 pA) representing weakly adapting and $w_\mathrm{jump}=0.0189$ (200 pA) strongly adapting \cite{Nicola2013bif}. The amount of heterogeneity increases from left to right with $\Delta_\eta=0$ representing homogeneous input currents. 
Strictly speaking, $\Delta_\eta=0$ contradicts the Lorentzian ansatz in the derivation of the mean-field model. However, the numerical continuation in both Fig.~\ref{fig:Effect_Delay_gsyn_hw_wjump_filled} and Fig.~\ref{fig:Effect_Delay_gsyn_hw_wjump_I_net_filled}  shows that there are no abrupt, qualitative changes when $\Delta_\eta$ grows from 0 to $10^{-4}$, except the replacement of blue subcritical Hopf bifurcations by red supercritical ones in some intervals of $D$. Combined with the perturbation analysis in Sec.~\ref{sec:pert_analysis}, it is reasonable to interpret the mean-field model with $\Delta_\eta=0$ as an approximation of the system with very weak heterogeneity. 
%
%
Additionally, when $g_\mathrm{syn}$ approaches zero, accounting for a population of neurons with significantly weak coupling, the system undergoes a splitting of the Hopf branch when $\Delta_\eta$ jumps down from 0.01 to $10^{-4}$, similar to the phenomenon shown in Fig.~\ref{fig:Effect_hw_wjump_NoDelay} (b) and (c), where the system has no synaptic delay. 

Now we concentrate on the bifurcation diagrams in the last two columns of Fig.~\ref{fig:Effect_Delay_gsyn_hw_wjump_filled}, where the input currents have moderate heterogeneity.
The grey areas show the regions of the parameters where EPs are stable, and thus neurons exhibit asynchronous tonic firing. 
For weak adaptation, $w_\mathrm{jump}=0.0095$, there are alternating intervals of $D$ where stable EPs exist. As $w_\mathrm{jump}$ grows, these intervals shrink and disappear, due to Hopf branches moving to the right towards the larger delay values, and new branches emerging from the $g_\mathrm{syn}$ axis.
At the same time, the blue curves representing subcritical Hopf bifurcation become shorter. 
Therefore, the adaptation favours the emergence of COs in the delayed excitatory neural network, demonstrated by the shrinking EP areas.
By comparison,  the increase of $\Delta_\eta$ from 0.01 to 0.02 expands EP areas as the Hopf curves move upwards, but also shortens the blue parts of the curves. So the heterogeneity promotes EP for small values of $g_\mathrm{syn}$.
Regarding the effect of synaptic delay $D$, for small enough synaptic coupling strength,  $g_\mathrm{syn}<0.5$ approximately, a stable EP area exists for all delay, $D>0$. For $g_\mathrm{syn}$ large enough, the delay generally promotes COs for strong adaptation ($w_\mathrm{jump}\ge 0.0189$) since the white areas exist for most values of $D$. As the adaptation weakens, e.g., $w_\mathrm{jump}=0.0095$, things become complicated: increasing the delay can both induce and destroy COs.
 

When it comes to the moderately heterogeneous current in the inhibitory network, as shown in the last two columns of 
Fig.~\ref{fig:Effect_Delay_gsyn_hw_wjump_I_net_filled},  one can see that the trends are the same as those in the excitatory network. The grey EP areas are independent of $D$ for small $g_\mathrm{syn}$.  The  Hopf branches move to right and up while the blue subcritical parts shrink with the increase of $w_\mathrm{jump}$ and $\Delta_\eta$. 
However, these changes yield different effects from the excitatory network because of the different bifurcation structure. Specifically, both $\Delta_\eta$ and $w_\mathrm{jump}$ do not favour the emergence of COs for $D <2$ or $g_\mathrm{syn}<0.5$, approximately.
In addition, the inhibitory coupling is more favourable for macroscopic oscillations than the excitatory one in the range of $D<2$ since the grey area is smaller than that in the excitatory case.
The white CO areas for the inhibitory and excitatory couplings are comparable for $D>2$, except the weak adaptation case $w_\mathrm{jump}=0.0095$.
More Hopf branches appear in the same range of $D$ for the excitatory coupling, yielding complicated oscillating dynamics.

\begin{figure}
\centering
\includegraphics[width=\textwidth]{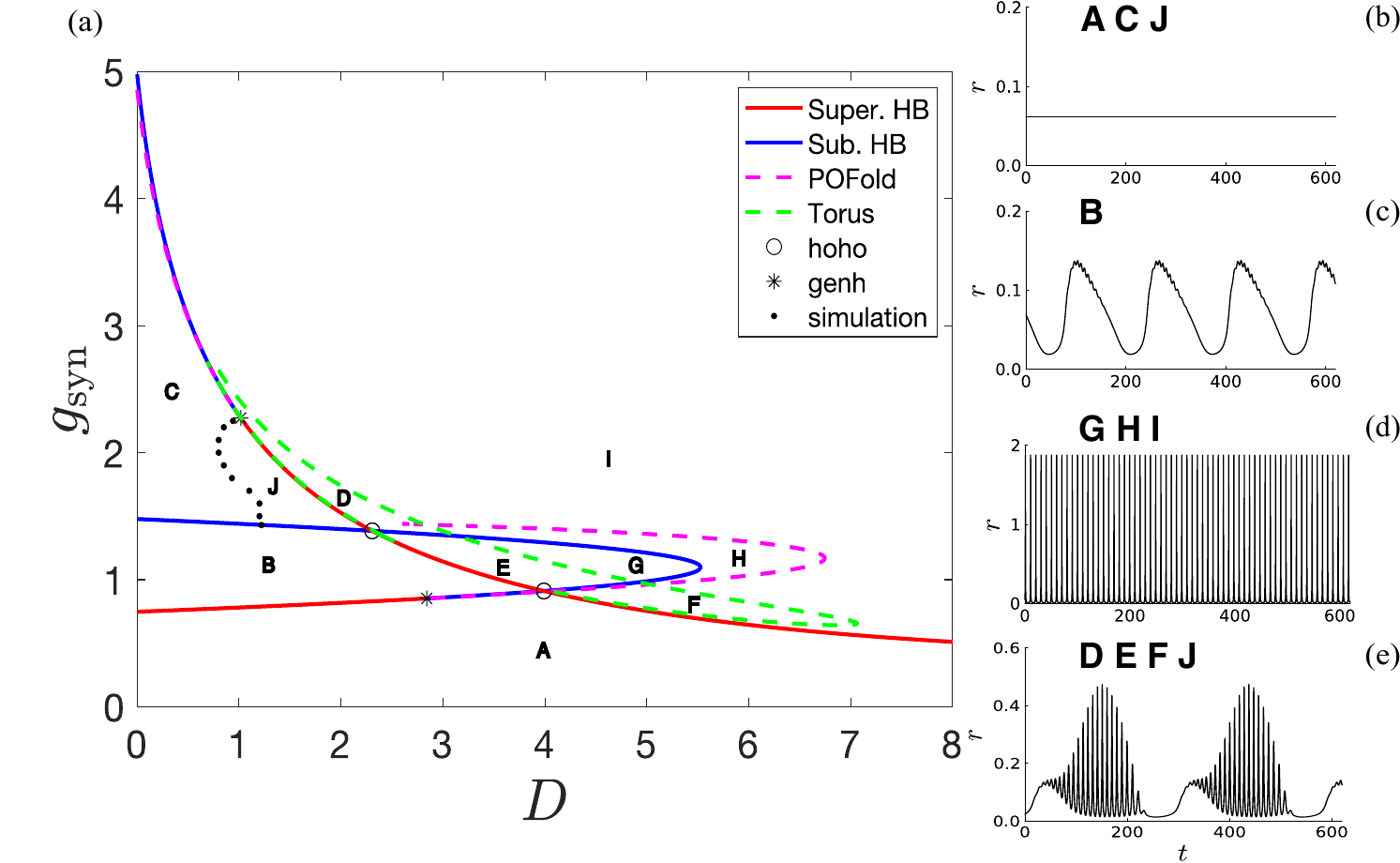}
\caption{Excitatory neural network with homogeneous synaptic delay.
(a) Bifurcation diagram in the $(D,g_\mathrm{syn})$ plane showing supercritical/subcritical Hopf bifurcations (solid red/blue lines), Fold bifurcation of periodic orbits (pink dash line) and Torus bifurcation (green dash line).  The symbols refer to codimension two bifurcation points: Hopf-Hopf (hoho) (circle), generalized Hopf (genh) (star).
The black dotted line, separating regions $C$ and $J$, has been determined by direct simulations of the mean-field model.
Sample time traces of $r(t)$ and $w(t)$ are shown in panels (b)-(d) for the three possible dynamical regimes:
(b) Region $A$, $C$, EP at $(D, g_\mathrm{syn})=(2,0.6)$,
(c) Region $B$, slow PO at $(D, g_\mathrm{syn})=(2,1)$,
(d) Region $G$, $H$, $I$, fast PO at $(D, g_\mathrm{syn})=(6,1.6)$,
(e) Region $D$, $E$, $F$, slow-fast nested COs at $(D, g_\mathrm{syn})=(4,1)$.
Region $J$: coexistence between EPs and slow-fast nested COs.
Other parameter values: $\bar \eta=0.25$, $\Delta_\eta=0.02$ and $w_\mathrm{jump}=0.025$.
}
\label{fig:TimeSeries_bif_E}
\end{figure}

Hopf bifurcations can only separate stationary solutions and oscillating states. However, different kinds of COs have been observed in the neural network, such as those shown in Fig.~\ref{fig:TimeSeries_RasterPlot}. To explore such rich macroscopic dynamics and understand the bifurcation mechanism underlying their emergence, further bifurcation analysis is necessary. 
For simplicity, we take two typical sets of parameters, one for the excitatory network and one for the inhibitory, and focus on bifurcations in a subset of the parameter space.
As shown in Fig.~\ref{fig:TimeSeries_bif_E} for the excitatory network, in addition to the Hopf-Hopf bifurcations (circle hoho points) and generalized Hopf bifurcations (star genh points) seen in  previous figures, we find two bifurcations of POs, thus creating different regions with distinct types of dynamics. One is the fold bifurcation of POs (pink dash line) originating from the generalized Hopf points and forming the boundaries where stable and unstable POs meet and disappear.
The other is the Torus (or Neimark-Sacker) bifurcation (green dash line) emanating from the Hopf-Hopf points and connecting with each other. The location of these bifurcation curves is consistent with the normal form coefficients of the two Hopf-Hopf points obtained by DDE-Biftool. For both points the system is in the subcase IV ($\theta>0$, $\delta<0$) of the difficult case (${\rm Re}(g2100)\cdot {\rm Re}(g0021)<0$) of Hopf-Hopf bifurcations (see \cite{Kuznetsov_BifBook}). Given the regions where the slow-fast nested COs occur, it is evident that the Torus bifurcation is the main mechanism for the emergence of these solutions. 
Samples of the time evolution of $r(t)$ are shown to the right of the bifurcation diagram, and they correspond to the following regimes:
\begin{itemize}
\item[(1)] Region $A$ and $C$: stable EPs
\item[(2)] Region $B$: stable POs with low frequency
\item[(3)] Region $G$, $H$ and $I$: stable POs with high frequency
\item[(4)] Region $D$, $E$ and $F$: slow-fast nested COs
\item[(5)] Region $J$: coexistence of stable EPs and slow-fast nested COs.
\end{itemize}
The black dotted line, separating
Region $J$ from Region $C$, was identified by direct simulations of the mean-field model.
Note that the PO exhibited in Fig.~\ref{fig:TimeSeries_bif_E} (c) is not strictly a regular spiking PO as that shown in Fig.~\ref{fig:TimeSeries_RasterPlot} (b). Small ripples appear in the time series of $r(t)$. However, it is typical behaviour of a system with multiple time scales and characterized by a limit cycle with a period related to the slow dynamics of the adaptation variable $w(t)$ and fast-damped oscillations on the amplitude resulting from the focus type of solutions of the fast subsystem involving $r(t)$, $v(t)$ and $s(t)$ variables. This can be explained by standard slow-fast analysis, e.g., \cite{Bertram2017}. 
Based on the bifurcation diagram in Fig.~\ref{fig:TimeSeries_bif_E} (a) and the smooth oscillations of $w(t)$ (see difference between  Fig.~\ref{fig:TimeSeries_RasterPlot}) (b) and (c), we don't categorize this behaviour as  slow-fast nested COs.

\begin{figure}
\centering
\includegraphics[width=\textwidth]{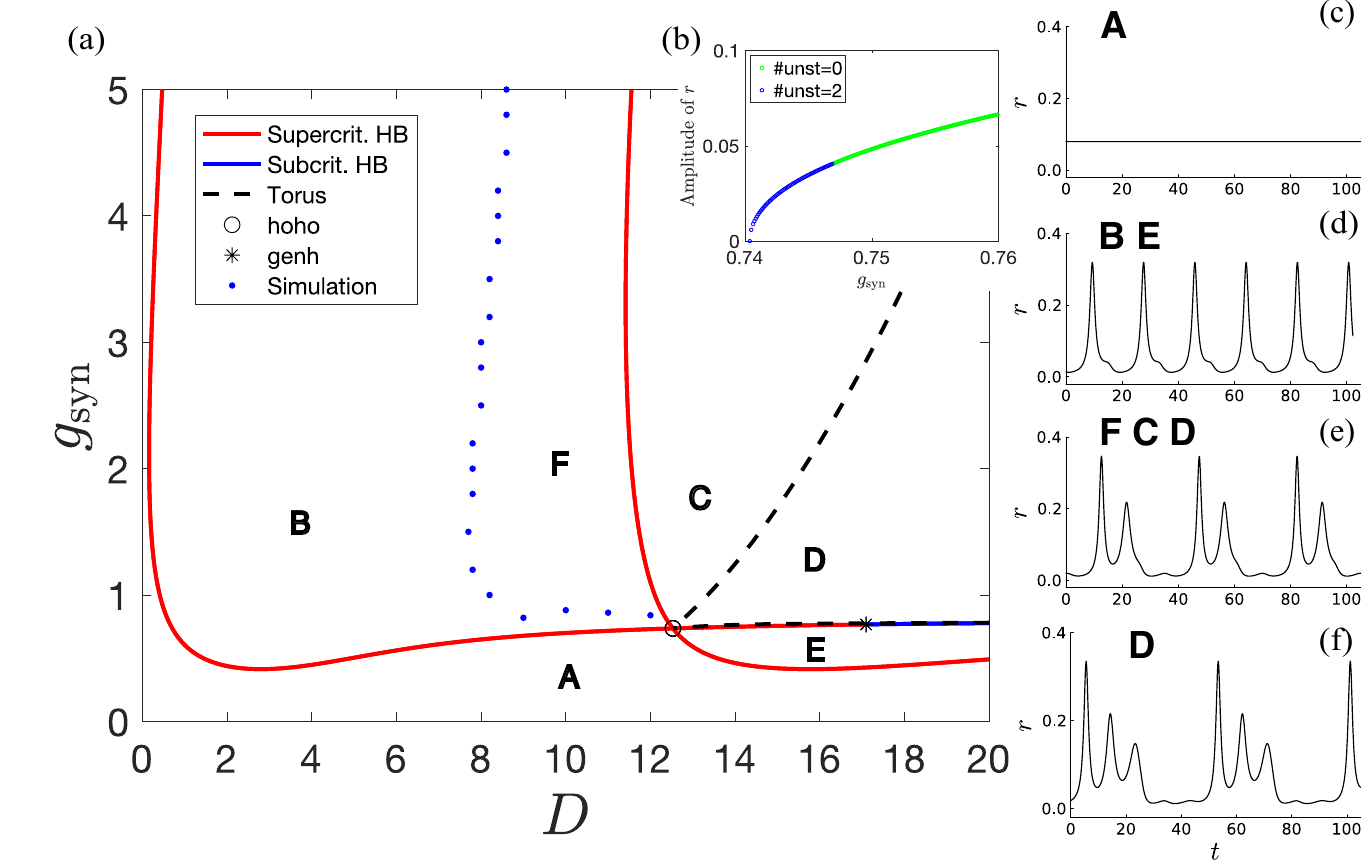}
\caption{Inhibitory neural network with homogeneous synaptic delay.
(a) Bifurcation diagram in the $(D,g_\mathrm{syn})$ plane showing supercritical/subcritical Hopf bifurcations (solid red/blue lines) and Torus bifurcations (black dash line).  The symbols refer to codimension two bifurcation points: Hopf-Hopf (hoho) (circle), generalized Hopf (genh) (star).
The blue dotted line, separating regions $B$ and $F$, has been determined by direct simulations of the mean-field model.
(b) Branch of periodic orbits emanating from close to the Hopf-Hopf point ($D=13$). Green/blue represents zero/two unstable Floquet multipliers.
Sample time traces of $r(t)$ are shown in panels (c)-(f) for the four possible dynamical regimes:
(c) Region $A$, EP at $(D, g_\mathrm{syn})=(6,0.4)$,
(d) Region $B$ and $E$, PO at $(D, g_\mathrm{syn})=(6,1)$,
(e) Region $F$, $C$ and $D$, double-period limit cycle at $(D, g_\mathrm{syn})=(14,1)$,
(f) Region $D$, three-period limit cycle at $(D, g_\mathrm{syn})=(20,1)$.
Other parameter values: $\bar \eta=0.4$, $\Delta_\eta=0.02$, $w_\mathrm{jump}=0.0189$.
}
\label{fig:TimeSeries_bif_I}
\end{figure}

By comparison, Fig.~\ref{fig:TimeSeries_bif_I} presents the detailed bifurcation diagram for the inhibitory network.
The system undergoes a Hopf-Hopf bifurcation (circle hoho point) and a generalized Hopf bifurcation (star genh point) as seen before. The normal form coefficients show that the system is in subcase I ($\theta>0$, $\delta>0$, $\theta\delta>1$)) of the simple case ((${\rm Re}(g2100)\cdot {\rm Re}(g0021)>0$) of the Hopf-Hopf bifurcation (see \cite{Kuznetsov_BifBook}). As predicted, there are two Torus bifurcation curves (black dash lines) emanating from the Hopf-Hopf point, see Fig.~\ref{fig:TimeSeries_bif_I} (a), and the Torus bifurcation curves lie between the two Hopf branches emanating from
the Hopf-Hopf point.
Sample time evolutions of $r(t)$ are shown in Fig.~\ref{fig:TimeSeries_bif_I} (c)-(f), and they correspond to the following regimes:
\begin{itemize}
\item[(1)] Region $A$: stable EPs
\item[(2)] Region $B$ and $E$: stable POs
\item[(3)] Region $F$, $C$: double-period limit cycles
\item[(4)] Region $D$: double-, triple-period limit cycles.
\end{itemize}

There are two discrepancies between the prediction of the bifurcation analysis and the time evolutions. 
First, the observable period-doubled limit cycles in Region $F$ indicate that the system undergoes a period-doubling bifurcation when $D$ crosses the blue dotted line. Unfortunately, we can only find the boundary by direct simulations of the mean-field model. It was not found by the numerical continuation. 
Second, according to the prediction of the normal form coefficients in \cite{Kuznetsov_BifBook}, there should be two stable periodic orbits between the Torus
bifurcation curves near the Hopf-Hopf point, that is, in Region $D$. The numerical continuation supports this conclusion, as shown in Fig.~\ref{fig:TimeSeries_bif_I} (b). A Torus bifurcation occurs when the number of unstable Floquet multipliers changes from 2 to 0, stabilizing the period orbit (green line). Unfortunately, our numerical simulations in Region $D$ found only one stable periodic orbit in this region, which is of double-, and even triple-period type.
In addition, we didn't see a cascade of period-doubling bifurcations, well known for their capability of inducing chaos, in our present choice of parameter values.
These interesting findings may arise from  other complexities in the bifurcation structure of our mean-field model. We note that the two frequencies of the POs near the Hopf-Hopf, (0.21 vs 0.53), are close to the ratio $1:2$. Period doubling bifurcations are known to occur in $1:2$ resonant Hopf-Hopf bifurcations \cite{leblanc1996classification}.  We leave further investigation of these and other phenomena for future work.


\section{Discussion}\label{sec:discussion}
We used numerical bifurcation analysis of a  mean-field reduction to study the effect of neural heterogeneity, spike frequency adaptation and time delayed connectivity on the emergence of coherent oscillations in  a large-scale network of Izhikevich-type  neurons.
The mean-field models were derived from the full network model using the theoretical framework of the Ott-Antonsen reduction \cite{Ott2008} via the equivalent Lorentzian ansatz \cite{Montbrio2015}.
Here we extended the specific mean-field approach for the neural network without delay \cite{ChenCampbell2022_mf}  to the network with distributed-delay coupling.
%
%
We performed perturbation analysis on the mean-field system with general distributed delay and focused the bifurcation analysis on the system with homogeneous delay.
Our study emphasized the effect of the heterogeneity of the input current $\Delta_\eta$, the adaptation intensity $w_\mathrm{jump}$ and the synaptic delay $D$ on the macroscopic dynamics.

Both perturbation and bifurcation analysis show that the mean-field model with $\Delta_\eta=0$ is a good approximation of the neural network with extremely weak heterogeneity. This is surprising as the Watanabe-Strogatz theory \cite{WS1993}, rather than the Ott-Antonsen theory used in this paper, is the correct framework to derive the mean-field model if the neurons are strictly identical. In fact, we see smooth variation of the bifurcation structure as $\Delta_\eta\rightarrow 0$, for $g_\mathrm{syn}$ sufficiently large. 
Our results provide evidence and a reasonable justification for some papers, e.g., \cite{Ratas2018_GammaDistr,PazoMontbrio2016,DevalleMontbrio2017}, where their analysis built on the Lorentzian-ansatz based mean-field model with $\Delta_\eta=0$. 
The mean-field model with weak heterogeneity may also be able to represent a homogeneous neural network under small random perturbations. The paper \cite{Montbrio2015} showed near quantitative agreement between the network of identical quadratic integrate-and-fire (QIF) neurons driven by Gaussian noise and the corresponding mean-field model derived under the Lorentzian ansatz. For our mean-field model derived from Izhikevich neurons,  we leave this for future investigation.  
To study the effects of $\Delta_\eta$ and $w_\mathrm{jump}$ on collective oscillations, we presented Hopf bifurcation diagrams in different parameter planes (see Fig.~\ref{fig:Effect_hw_wjump_NoDelay}-\ref{fig:Effect_Delay_gsyn_hw_wjump_I_net_filled}). Essentially, all these results follow one principle, that is, the emergence of collective oscillations is due to a balance between the excitatory drives, which cause neurons to spike and the inhibitory drives, which terminate spiking. The intrinsic and external input currents $\eta$ and $I_\mathrm{ext}$ are typical excitatory drives, and the adaptation current $w$ is the inhibitory drive.
The role of the synaptic current $I_\mathrm{syn}$ depends on the sign of $e_r-v_k$, which is negative in the inhibitory network and positive/negative in the excitatory case. 
The synaptic delay does not affect the number and the position of the equilibrium points but does influence their stability. Generally, for the present choice of parameter values, delays have little influence on generation of COs when $g_\mathrm{syn}$ is small, which represents the weak coupling (see Fig.~\ref{fig:Effect_Delay_gsyn_hw_wjump_filled} and \ref{fig:Effect_Delay_gsyn_hw_wjump_I_net_filled}). Beyond that range, delays work like an excitatory drive to promote the emergence
of COs and generate new dynamics, including slow-fast nested COs resulting from Torus bifurcations (see Fig.~\ref{fig:TimeSeries_bif_E}).

There are many papers in the literature studying the neural network with delayed coupling. To our knowledge, three papers \cite{Ratas2018_GammaDistr,PazoMontbrio2016,DevalleMontbrio2017} are closely related to ours. 
These papers all used the same theoretical framework to derive the mean-field model as we do, and used the model to study the macroscopic dynamics induced by the coupling delay. 
However, in this paper, we derived mean-field models from neural networks with more neurological plausibility and performed more detailed bifurcation analysis.
First, we took into account the spike frequency adaptation in each neuron. This is a fundamental neuronal mechanism and plays a prominent role in promoting synchronous bursting of the neurons. Second, we employed a realistic expression for the synaptic current \eqref{eq:synaptic_current} in the network connections. This model is widely used in neuroscience, but often simplified for analysis. This simplification loses the physiological insight and may lead to discrepancies in the prediction of system behaviours.
The increased biological relevance of our neural model leads to a more complicated mean-field model. It is four dimensional due to the addition of the adaptation variable and contains new nonlinear terms arising from the more complicated synapse model. Thus, analytical Hopf bifurcation analysis such as in \cite{PazoMontbrio2016, DevalleMontbrio2018_delay, Ratas2018_GammaDistr} is not tractable. Instead, we used numerical approaches. This has the added benefit of allowing us
to locate higher-codimension bifurcations and study their effect on the macroscopic dynamics.
%

Our mean-field system exhibits significantly different Hopf bifurcation structures than those observed in \cite{PazoMontbrio2016, DevalleMontbrio2018_delay}. Many Bautin/generalized Hopf bifurcation points occur in the excitatory network. These are points where the Hopf bifurcation switches between supercritical and subcritical. Also, Hopf-Hopf bifurcations occur in both excitatory and inhibitory networks.
Further bifurcation analysis of these new structures allowed us to discover Fold bifurcations of periodic orbits, where stable and unstable POs meet and disappear, and Torus bifurcations. In the excitatory network, the Torus bifurcations lead to the creation of a stable torus solution, corresponding to  
slow-fast nested COs. In the inhibitory network the numerical bifurcation analysis found a different arrangement of the Torus bifurcations, leading to bistability between two different POs. Using direct numerical simulation, however, we did not observe the bistability. Instead we found evidence for a period-doubling bifurcation. Interestingly, we didn't find a cascade of period-doubling bifurcations leading to chaos, as was observed in the inhibitory network in \cite{PazoMontbrio2016,Ratas2018_GammaDistr}.
The discrepancy between the numerical continuation and numerical simulation in the inhibitory network may be due to a further complication in the system: the frequencies at the Hopf-Hopf points we investigated are close to 1:2 resonance. Such resonant Hopf-Hopf points can lead to isolated period doubling bifurcations~\cite{leblanc1996classification}. We will leave this for further investigation in the future. 

The slow-fast COs emerging from Torus bifurcations in the excitatory network are different than those observed in some other papers, which typically consist of a small amplitude fast "ripple" riding on a large amplitude slow oscillation. In this paper, the fast oscillations may have large amplitude and the slow oscillations modulate this amplitude, similar as the recordings observed in the Hippocampus and various brain regions \cite{lisman2008_CFC}. 
%
Other papers have shown that the fast oscillations can be made to fall in the $\gamma$ range (30-100 Hz) and the slow modulation in the $\theta$ band (4-8 Hz) by choosing appropriate parameter values, e.g., the adaptation intensity $w_\mathrm{jump}$ \cite{Ferrara2023}. In our COs, the slow oscillations are generated by the adaptation, while fast oscillations are due to the intrinsic spiking frequency of the neurons and the time delay. For the parameters of Fig.~\ref{fig:TimeSeries_RasterPlot} (c), the slow oscillations correspond to ~$2$ Hz and the fast to $70$ Hz when converted to dimensional values. These two values can be modified by adjusting the adaptation intensity and the size of the time delay, respectively.
$\theta$-nested $\gamma$ oscillations have been observed in the hippocampus and shown to be relevant to cognitive tasks, such as navigation, sensory association and working memory \cite{Belluscio2012, Butler2016, Buszaki2012, Colgin2009_nature, lisman2008_CFC}. 
Generally, this interaction between different frequency bands belongs to a phenomenon termed Cross-Frequency-Coupling (CFC), which are potentially relevant for understanding healthy and pathological brain functions \cite{canolty2010_CFC,aru2015_CFC, lisman2008_CFC}.
The emergence of CFC has been reported for next-generation neural mass models, e.g., for two coupled neural networks \cite{ceni2020_CFC,Ferrara2023}, or  for a network with an external $\theta-$drive \cite{segneri2020_CFC}. By contrast, we have shown that CFC may emerge in a single population of neurons with spike frequency adaptation and delayed excitatory coupling.
%

\section*{ACKNOWLEDGMENTS}
This work benefited from the support of the Natural Science and Engineering Research Council of Canada.


\appendix
\setcounter{figure}{0} 
\setcounter{table}{0} 


\section{Using perturbation analysis to find equilibria}\label{app:pert_EP}

We apply a perturbation method to find the equilibria of the general mean-field model \eqref{eq:single-exp-synapse_s_u}, \eqref{eq:mf_izh_rvw} and \eqref{eq:spike_train_distr_delay} when the input current $\eta$ has a weak heterogeneous distribution, that is, $0< \Delta_\eta <<  1$.

Standard analysis shows that the equilibria obey the following equations,
\begin{equation*}
\label{eq:mf_EP}
\begin{split}
\Delta_\eta/\pi 
+
2 r_s v_s 
-
\big(\alpha + g_{\mathrm{syn}} s_s \big) r_s
&= 0
\\
v_s^2
- \alpha v_s
-
\pi^2 r_s^2
+
g_{\mathrm{syn}} s_s 
\big(
e_r - v_s 
\big)
-
w_s 
+
\bar \eta + I_{\mathrm{ext}}
&=0
\\
a
\left (
b 
v_s 
-
w_s 
\right )
+
w_{\mathrm{jump}} r_s
&= 0
\\
-s_s/\tau_s
+
s_{\mathrm{jump}} r_s
&= 0 
\end{split}
\end{equation*}
Here, we employ
\begin{equation*}
\int_0^\infty h(\tau)d\tau = 1.
\end{equation*}
After some calculations, we have
\begin{equation*}
\begin{split}
v_s
&=
\frac{J}{2} r_s
-
\frac{\Delta_\eta}{2\pi}\frac{1}{r_s}
+
\frac{\alpha}{2}
\\
w_s
&=
b v_s 
+
\frac{w_\mathrm{jump}}{a}r_s
\\
s_s
&=
\tau_s s_\mathrm{jump} r_s
\end{split}
\end{equation*}
where $J=g_\mathrm{syn}\tau_s s_\mathrm{jump}$
and $r_s$ follows the polynomial equation 
\begin{equation*}
C_4 r_s^4 
+
C_3 r_s^3
+
C_2 r_s^2
+
C_1 r_s
+
C_0
=0,
\end{equation*}
with
\begin{equation*}
\begin{split}
C_4 
&=
J^2 + 4\pi^2,
\\
C_3 
&= 
2J 
(\alpha + b -2e_r)
+
4 w_\mathrm{jump}/a,
\\
C_2 
&= 
\alpha^2 + 2\alpha b - 4I_\mathrm{ext} -4\bar \eta,
\\
C_1 
&=
-2 b \Delta_\eta /\pi
\\
C_0
&=
-\Delta_\eta^2/\pi^2.
\end{split}
\end{equation*}

Since the heterogeneity is weak, we suppose that $\Delta_\eta$ is a small parameter, and define $\epsilon=\Delta_\eta <<1$. 
Then, we rewrite the equation for $r_s$ as
\begin{equation*}
C_4 r_s^4
+
C_3 r_s^3
+
C_2 r_s^2
+
\widetilde C_1\epsilon r_s
+
\widetilde C_0\epsilon^2 =0    
\end{equation*}
where $\widetilde C_1=-2b/\pi$ and $\widetilde C_0=-1/\pi^2$. 
We then look for roots of the above equation in the form
\begin{equation*}
r_s
=
r_{s,0}
+r_{s,1}\epsilon
+r_{s,2}\epsilon^2
+O(\epsilon^3)
\end{equation*}
Substituting it, expanding and collecting terms in like powers of $\epsilon$ we obtain the $O(1)$ equation
\begin{equation*}
C_4 r_{s,0}^4
+
C_3 r_{s,0}^3
+
C_2 r_{s,0}^2=0  
\end{equation*}
This has the solutions $r_{s,0}=0,0$ and 
\begin{equation}
\label{eq:mf_ep_pert_nonzero}
r_{s,0}^\pm
=
\frac{- C_3 
\pm
\sqrt{ C_3^2 - 4 C_4 C_2}}
{2 C_4}
\end{equation}

The $O(\epsilon)$ equation is 
\begin{equation*}
4C_4 r_{s,0}^3 r_{s,1}
+
3C_3 r_{s,0}^2 r_{s,1}
+
2C_2 r_{s,0} r_{s,1}
+
\widetilde C_1 r_{s,0}
=0   
\end{equation*}
With $r_{s,0}=0$, this gives no constraint on $r_{s,1}$. 
With $r_{s,0} \neq 0$, we have
\begin{equation*}
r_{s,1}^\pm
=
\frac{-\widetilde C_1}
{4 C_4 r_{s,0}^2 
+
3 C_3 r_{s,0}
+
2 C_2},
\quad \mathrm{when} \quad r_{s,0} \neq 0
\end{equation*}

The $O(\epsilon^2)$  equation is
\begin{equation*}
\big (
4C_4 r_{s,0}^3
+
3C_3 r_{s,0}^2
+
2C_2r_{s,0}
\big )r_{s,2}
+
\big (
6C_4 r_{s,0}^2
+
3C_3r_{s,0}
+
C_2
\big )r_{s,1}^2
+
\widetilde C_1 r_{s,1}+\widetilde C_0
=0
\end{equation*}
With $r_{s,0}=0$, this gives no constraint on $r_{s,2}$. 
However, $r_{s,1}$ should satisfy 
\begin{equation*}
\big (
6C_4 r_{s,0}^2
+
3C_3r_{s,0}
+
C_2
\big )r_{s,1}^2
+
\widetilde C_1 r_{s,1}+\widetilde C_0
=0
\end{equation*}
which gives 
\begin{equation*}
r_{s,1}^{\pm}
=
\frac{- \widetilde C_1
\pm
\sqrt{\widetilde C_1^2
-
4 \widetilde C_0 C_2}
}
{2C_2},
\quad \mathrm{when} \quad r_{s,0} = 0
\end{equation*}
Thus, the system has EPs, to the lowest order, at 
\begin{equation}
\label{eq:mf_ep_pert_zero}
r_s
=
r_{s,1}^\pm \cdot \epsilon
=
\frac{- \widetilde C_1
\pm
\sqrt{\widetilde C_1^2
-
4 \widetilde C_0 C_2}
}
{2C_2}
\cdot \epsilon
\quad \mathrm{when} \quad r_{s,0} = 0
\end{equation}
Since $0<\epsilon <<1$, these solutions are very close to zero.
With $r_{s,0} \neq 0$, we have
\begin{equation*}
\label{eq:mf_ep_r2}
r_{s,2}^\pm
=
-\frac{
\big (
6C_4r_{s,0}^2
+
3C_3r_{s,0}
+
C_2
\big )r_{s,1}^2
+
\widetilde C_1r_{s,1}+\widetilde C_0
}
{
4C_4r_{s,0}^3
+
3C_3r_{s,0}^2
+
2C_2r_{s,0}
},
\quad \mathrm{when} \quad r_{s,0} \neq 0
\end{equation*}
%
In summary, the mean-field system \eqref{eq:single-exp-synapse_s_u}, \eqref{eq:mf_izh_rvw} and \eqref{eq:spike_train_distr_delay} with weak heterogeneity of the current $\eta$, has  at most four EPs, given to lowest order by \eqref{eq:mf_ep_pert_nonzero} and \eqref{eq:mf_ep_pert_zero} and the latter two are close to $r_s=0$.
Note that the expressions for the EPs are well behaved as $\epsilon\rightarrow 0$. Thus the EPs in the homogeneous system correspond to the $O(1)$ solutions above.

Next, we can obtain the region in the parameter space for the existence of $r_s$. See Fig.~\ref{fig:EP_domain_E_pert} (a) for reference.
From Eq.~\eqref{eq:mf_ep_pert_nonzero}, we have 
\begin{itemize}
\label{list:ep_para_space}
\item Only $r_s=r_{s,0}^+ >0$ exists when $C_2 < 0$, corresponding to the right of the blue line, or $C_2=0$ and $C_3<0$, the blue solid line below the green dashed line.
\item Both $r_s=r_{s,0}^\pm >0$ exist when $0<C_2<\frac{C_3^2}{4C_4}$ (bounded by the red line and the blue dashed line) and $C_3<0$ (above the green dashed line). Both solutions are born at a saddle-node bifurcation when $C_2=\frac{C_3^2}{4C_4}$.
\end{itemize}
Here, $C_4>0 $ is taken into account, also note that the value of $\bar \eta$ changes the sign of $C_2$, and $J$ changes the sign of $C_3$. 
Similarly, we can obtain the domain for the existence of two solutions close to zero from Eq.~\eqref{eq:mf_ep_pert_zero}.
Note that $\widetilde C_0<0$ and $\mathrm{sgn}(\widetilde C_1)=-\mathrm{sgn}(b)$. Then, we have
\begin{itemize}
\item Only $r_s = r_{s,1}^+\epsilon>0$ exists when $C_2>0$, corresponding to the left of the blue line in Fig.~\ref{fig:EP_domain_E_pert} (a). It works for both $b<0$ and $b>0$.
\item Both $r_s=r_{s,1}^{\pm}\epsilon>0$ exist when $\frac{\widetilde C_1^2}{4 \widetilde C_0} < C_2 <0$ and $b<0$. 
\end{itemize}
However, $\frac{\widetilde C_1^2}{4 \widetilde C_0} < C_2 <0$ gives
\begin{equation*}
\frac{\alpha^2+2\alpha b}{4}
<
I_\mathrm{ext} + \bar \eta 
<  
\frac{\alpha^2+2\alpha b}{4}
+
\frac{b^2}{4},
\end{equation*}
where $b=-0.0062$ in \cite{ChenCampbell2022_mf} is so small that the region for the existence of $r_s=r_{s,1}^{\pm}\epsilon>0$ is extremely narrow such that we can neglect it.

If we consider the inhibitory neural network where $e_r$ is negative, we have $C_3>0$ and $C_4>0$. So, on the right of the blue line in Fig.~\ref{fig:EP_domain_E_pert} (a), only $r_s=r_{s,0}^+>0$ exists and on the left of the blue line, only $r_s=r_{s,1}^+\epsilon \approx 0$ exists.


\section{Characteristic equation}
\label{app:char_EP_HP}
In this section, we show how to derive
the characteristic equation of the linearized mean-field system and parametric equations for the Andronov-Hopf bifurcation.

We substitute $r = r_s + \delta r$,
$v = v_s + \delta v$,
$w = w_s + \delta w$ and $s = s_s + \delta s$ into the mean-field model with generally distributed delays Eqs.~\eqref{eq:single-exp-synapse_s_u}, \eqref{eq:mf_izh_rvw} and \eqref{eq:spike_train_distr_delay} and linearize them with respect to small deviations $(\delta r, \delta v, \delta w, \delta s)$. Then we get the linearized equations
\begin{equation*}
\label{eq:mf_linear_distr}
    \begin{split}
        \delta r' (t)
        & =
        (2v_s - \alpha - g_\mathrm{syn} s_s) \delta r
        +
        2 r_s \delta v
        -
        g_\mathrm{syn} r_s \delta s
        \\
        \delta v'(t)
        &=
        -2\pi^2 r_s \delta r
        +
        (2 v_s - \alpha - g_\mathrm{syn} s_s) \delta v
        -
        \delta w
        +
        g_\mathrm{syn}(e_r -  v_s) \delta s
        \\
        \delta w'(t)
        &=
        w_\mathrm{jump} \delta r
        + 
        ab \delta v
        -
        a \delta w
        \\
        \delta s'(t)
        &=
        - \delta s /\tau_s 
        +
        s_\mathrm{jump}
        \int_0^\infty
        \delta r(t - \tau)
        h(\tau)
        d \tau
    \end{split}
\end{equation*}

We look for the solution of the linearized equations in the form 
\begin{equation*}
(\delta r, \delta v, \delta w, \delta s) =(K_r,K_v,K_w,K_s)\exp(\lambda t),    
\end{equation*}
where $K_j$ are constants. Then, we have
\[ [{\bf A}(\lambda)-\lambda \bf{I}]\,{\bf K}={\bf 0}\]
where ${\bf K}=(K_r,K_v,K_w,K_s)^T$ and 
\[ {\bf A}(\lambda)
=
\begin{bmatrix}
2v_s-\alpha-g_\mathrm{syn} s_s 
& 
2r_s
&
0
&
-g_\mathrm{syn}r_s
\\
-2\pi^2 r_s
&
2v_s-\alpha-g_\mathrm{syn}s_s
&
-1
&
g_\mathrm{syn}(e_r - v_s)
\\
w_\mathrm{jump}
&
ab
& 
-a
&
0
\\
s_\mathrm{jump}H(\lambda) & 0 &0 & -1/\tau_s
\end{bmatrix}
\]
where $H(\cdot)$ is the Laplace transform of $h(\cdot)$, that is,
\begin{equation*}
H(\lambda)
=
\int_0^\infty
e^{-\lambda \tau}
h(\tau)d\tau.
\end{equation*}

Then, we derive the characteristic equation 
\begin{equation*}
\Lambda(\lambda)
=
P(\lambda)H^{-1}(\lambda)
+
Q(\lambda)
\end{equation*}
where 
\begin{equation*}
\begin{split}
P(\lambda)
&=
\Big [
(\lambda + a)(\lambda + K)^2 
+
ab (\lambda + K)
+ 
4\pi^2 r_s^2 (\lambda + a)
+
2  w_\mathrm{jump} r_s
\Big ]
(\tau_s\lambda + 1) 
\\
Q(\lambda)
&=
\Big [
J r_s 
(\lambda + K)(\lambda + a)
-
2 J r_s   
(e_r - v_s)
(\lambda + a)
+ 
ab J r_s  
\Big ]
\end{split}
\end{equation*}
Here, we denote $J = \tau_s g_\mathrm{syn} s_\mathrm{jump}$ and $K = \alpha + J r_s - 2 v_s$.

Next, derive parametric equations for the Hopf bifurcation curves. Substituting $\lambda = i\omega$, $K = \alpha + J r_s - 2 v_s$ into the characteristic equation and collecting terms in like powers of $r_s$, we have
\begin{equation*}
\begin{split}
P(i\omega)
&=
R_2 r_s^2 
+
R_1 r_s 
+
R_0 +
i \omega \left(  
D_2 r_s^2 
+
D_1 r_s 
+
D_0
\right ) 
\\
Q(i\omega)
&=
\hat R_2 r_s^2 
+
\hat R_1 r_s
+
i\omega \left (
\hat D_2  r_s^2 
+
\hat D_1 r_s
\right ) 
\end{split}
\end{equation*}
where
\begin{equation*}
\begin{split}
R_2
&=
-
\tau_s 
(J^2 + 4\pi^2)
\omega^2
+
a J^2 + 4a \pi^2
\\
R_1
&=
-2 J
\left [
1+\tau_s (a+\alpha - 2 v_s)
\right ]
\omega^2
+
2aJ(\alpha - 2v_s)
+
a b J + 2w_\mathrm{jump}
\\
R_0
&=
\tau_s \omega^4
-
\big [
\tau_s(\alpha-2 v_s)^2
+
2(\alpha-2 v_s)
(\tau_s a + 1)
+
\tau_s ab
+a
\big ] 
\omega^2 \cdots
\\
& \qquad \qquad \qquad \qquad \qquad \qquad \qquad \qquad \qquad 
+a(\alpha-2 v_s)
(\alpha - 2 v_s +b)
\\
D_2
&=
(J^2+4\pi^2)(1 + a\tau_s)
\\
D_1
&=
-2J\tau_s \omega^2
+
2J(a+\alpha-2 v_s)
+
\tau_s
\big [
2aJ(\alpha - 2 v_s)
+
abJ
+
2w_\mathrm{jump}
\big ]
\\
D_0
&=
-\big [
1+\tau_s(a+2\alpha - 4 v_s)
\big ]
\omega^2
+
(\alpha - 2 v_s)^2
+ 2a(\alpha - 2v_s)
+ ab \cdots
\\
& \qquad \qquad \qquad \qquad \qquad \qquad \qquad \qquad
+
a\tau_s
\big [
(\alpha - 2v_s)
(\alpha - 2v_s + b)
\big ]
\end{split}
\end{equation*}
and 
\begin{equation*}
\begin{split}
\hat R_2
&=
a J^2
\\
\hat R_1
&=
-J \omega^2
+ a J(\alpha + b - 2e_r)
\\
\hat D_2
&=
J^2 
\\
\hat D_1
&=
J(a + \alpha - 2e_r)
\end{split}
\end{equation*}
Denote $H^{-1}(i\omega) = A(\omega) + i B(\omega)$.  Separating the real and imaginary parts of the characteristic equation yields the parametric equations for the Hopf bifurcation curves,
\begin{equation*}
\begin{split}
(R_2 r_s^2 + R_1 r_s + R_0)A
-
(D_2 r_s^2 + D_1 r_s + D_0)B\omega
+
(\hat R_2 r_s^2 + \hat R_1 r_s)
&= 0
\\
(R_2 r_s^2 + R_1 r_s + R_0)B
+
(D_2 r_s^2 + D_1 r_s + D_0)A\omega
+
(\hat D_2 r_s^2 + \hat D_1 r_s)\omega
&= 0
\end{split}
\end{equation*}

A proper choice of the distribution function for the synaptic delay can further simplify the stability analysis.  
The simplest case is the neural network without synaptic delay, where $h(\tau) = \delta(\tau)$ is the Dirac delta function and the reciprocal of the Laplace transform at the Hopf point is $H^{-1}(j\omega) = 1$.
For the network with homogeneous delay $D$,  the distribution function is $h(\tau) = \delta(\tau - D)$ and the reciprocal of the Laplace transform is given by
\begin{equation*}
H^{-1}(i\omega) 
= e^{i\omega D}
= \cos(\omega D)
+ i\sin(\omega D)
\equiv A(D,\omega) + i B(D,\omega).
\end{equation*}



\bibliographystyle{elsarticle-num}
\bibliography{MFrefs}

\begin{thebibliography}{10}
\expandafter\ifx\csname url\endcsname\relax
  \def\url#1{\texttt{#1}}\fi
\expandafter\ifx\csname urlprefix\endcsname\relax\def\urlprefix{URL }\fi
\expandafter\ifx\csname href\endcsname\relax
  \def\href#1#2{#2} \def\path#1{#1}\fi

\bibitem{WilsonCowan1972}
H.~R. Wilson, J.~D. Cowan, Excitatory and inhibitory interactions in localized populations of model neurons, Biophysical Journal 12~(1) (1972) 1--24.
\newblock \href {http://dx.doi.org/10.1016/S0006-3495(72)86068-5} {\path{doi:10.1016/S0006-3495(72)86068-5}}.

\bibitem{WilsonCowan2021}
H.~R. Wilson, J.~D. Cowan, Evolution of the {W}ilson–{C}owan equations, Biological cybernetics 115~(6) (2021) 643--653.
\newblock \href {http://dx.doi.org/10.1007/s00422-021-00912-7} {\path{doi:10.1007/s00422-021-00912-7}}.

\bibitem{Ott2008}
E.~Ott, T.~M. Antonsen, Low dimensional behavior of large systems of globally coupled oscillators, Chaos 18~(3) (2008) 037113.
\newblock \href {http://dx.doi.org/https://doi.org/10.1063/1.2930766} {\path{doi:https://doi.org/10.1063/1.2930766}}.

\bibitem{WS1993}
S.~Watanabe, S.~H. Strogatz, Integrability of a globally coupled oscillator array, Physical Review Letters 70~(16) (1993) 2391--2394.
\newblock \href {http://dx.doi.org/10.1103/PhysRevLett.70.2391} {\path{doi:10.1103/PhysRevLett.70.2391}}.

\bibitem{Byrne2020}
{\'A}.~Byrne, R.~D. O'Dea, M.~Forrester, J.~Ross, S.~Coombes, Next-generation neural mass and field modeling, Journal of Neurophysiology 123~(2) (2020) 726--742.
\newblock \href {http://dx.doi.org/10.1152/jn.00406.2019} {\path{doi:10.1152/jn.00406.2019}}.

\bibitem{Ashwin2016Rew}
P.~Ashwin, S.~Coombes, R.~Nicks, Mathematical frameworks for oscillatory network dynamics in neuroscience, Journal of Mathematical Neuroscience 6~(1) (2016) 1--92.
\newblock \href {http://dx.doi.org/10.1186/s13408-015-0033-6} {\path{doi:10.1186/s13408-015-0033-6}}.

\bibitem{Bick2020}
C.~Bick, M.~Goodfellow, C.~R. Laing, E.~A. Martens, Understanding the dynamics of biological and neural oscillator networks through exact mean-field reductions: a review, Journal of Mathematical Neuroscience 10 (2020) 9.
\newblock \href {http://dx.doi.org/10.1186/s13408-020-00086-9} {\path{doi:10.1186/s13408-020-00086-9}}.

\bibitem{Katz1965_SynapticDelay}
B.~Katz, R.~Miledi, The measurement of synaptic delay, and the time course of acetylcholine release at the neuromuscular junction, Proceedings of the Royal Society of London. Series B, Biological sciences 161~(985) (1965) 483--495.
\newblock \href {http://dx.doi.org/10.1098/rspb.1965.0016} {\path{doi:10.1098/rspb.1965.0016}}.

\bibitem{Smith2011}
H.~Smith, Distributed delay equations and the linear chain trick, in: An Introduction to Delay Differential Equations with Applications to the Life Sciences, Springer, New York, NY, 2011, pp. 119--130.
\newblock \href {http://dx.doi.org/10.1007/978-1-4419-7646-8_7} {\path{doi:10.1007/978-1-4419-7646-8_7}}.

\bibitem{reddy1998time}
D.~R. Reddy, A.~Sen, G.~L. Johnston, Time delay induced death in coupled limit cycle oscillators, Physical Review Letters 80~(23) (1998) 5109.

\bibitem{Coombes1997_DelayPhaseLock}
S.~Coombes, G.~J. Lord, Intrinsic modulation of pulse-coupled integrate-and-fire neurons, Phys. Rev. E 56 (1997) 5809--5818.
\newblock \href {http://dx.doi.org/10.1103/PhysRevE.56.5809} {\path{doi:10.1103/PhysRevE.56.5809}}.

\bibitem{ShayerCampbell2000_DelayMultistab}
L.~P. Shayer, S.~A. Campbell, Stability, bifurcation, and multistability in a system of two coupled neurons with multiple time delays, SIAM Journal on Applied Mathematics 61~(2) (2000) 673--700.
\newblock \href {http://dx.doi.org/10.1137/S0036139998344015} {\path{doi:10.1137/S0036139998344015}}.

\bibitem{ChenCampbell2021_hysteresis}
L.~Chen, S.~A. Campbell, Hysteresis bifurcation and application to delayed {F}itz{H}ugh{N}agumo neural systems, Journal of Mathematical Analysis and Applications 500~(1) (2021) 125151.
\newblock \href {http://dx.doi.org/10.1016/j.jmaa.2021.125151} {\path{doi:10.1016/j.jmaa.2021.125151}}.

\bibitem{campbell2007_delay}
S.~A. Campbell, Time delays in neural systems, in: V.~K. Jirsa, A.~{McIntosh} (Eds.), Handbook of Brain Connectivity, Springer Berlin Heidelberg, 2007, pp. 65--90.
\newblock \href {http://dx.doi.org/10.1007/978-3-540-71512-2_2} {\path{doi:10.1007/978-3-540-71512-2_2}}.

\bibitem{wu2011introduction}
J.~Wu, Introduction to neural dynamics and signal transmission delay, Vol.~6, Walter de Gruyter, 2011.
\newblock \href {http://dx.doi.org/10.1515/9783110879971} {\path{doi:10.1515/9783110879971}}.

\bibitem{wu2022time}
J.~Wu, S.~A. Campbell, J.~B{\'e}lair, Time-delayed neural networks: Stability and oscillations, in: Encyclopedia of Computational Neuroscience, Springer, 2022, pp. 3434--3440.
\newblock \href {http://dx.doi.org/10.1007/978-1-4614-7320-6_513-2} {\path{doi:10.1007/978-1-4614-7320-6_513-2}}.

\bibitem{Ermentrout2009Delay}
B.~Ermentrout, T.-W. Ko, Delays and weakly coupled neuronal oscillators, Philosophical Transactions of the Royal Society A: Mathematical, Physical and Engineering Sciences 367~(1891) (2009) 1097--1115.
\newblock \href {http://dx.doi.org/10.1098/rsta.2008.0259} {\path{doi:10.1098/rsta.2008.0259}}.

\bibitem{Montbrio2015}
E.~Montbri{\'o}, D.~Paz{\'o}, A.~Roxin, Macroscopic description for networks of spiking neurons, Physical Review X 5 (2015) 021028.
\newblock \href {http://dx.doi.org/10.1103/PhysRevX.5.021028} {\path{doi:10.1103/PhysRevX.5.021028}}.

\bibitem{Izhikevich2003}
E.~M. Izhikevich, Simple model of spiking neurons, IEEE Transactions on Neural Networks 14~(6) (2003) 1569--1572.
\newblock \href {http://dx.doi.org/10.1109/TNN.2003.820440} {\path{doi:10.1109/TNN.2003.820440}}.

\bibitem{Ratas2018_GammaDistr}
I.~Ratas, K.~Pyragas, Macroscopic oscillations of a quadratic integrate-and-fire neuron network with global distributed-delay coupling, Physical Review E 98~(5) (2018) 052224.
\newblock \href {http://dx.doi.org/10.1103/PhysRevE.98.052224} {\path{doi:10.1103/PhysRevE.98.052224}}.

\bibitem{Ferrara2023}
A.~Ferrara, D.~Angulo-Garcia, A.~Torcini, S.~Olmi, Population spiking and bursting in next-generation neural masses with spike-frequency adaptation, Phys. Rev. E 107 (2023) 024311.
\newblock \href {http://dx.doi.org/10.1103/PhysRevE.107.024311} {\path{doi:10.1103/PhysRevE.107.024311}}.

\bibitem{DDE-BIFTOOL}
K.~Engelborghs, T.~Luzyanina, D.~Roose, Numerical bifurcation analysis of delay differential equations using dde-biftool, ACM Transactions on Mathematical Software 28~(1) (2002) 1--21.
\newblock \href {http://dx.doi.org/10.1145/513001.513002} {\path{doi:10.1145/513001.513002}}.

\bibitem{XPPAUT}
B.~Ermentrout, Simulating, Analyzing, and Animating Dynamical Systems: A Guide to XPPAUT for Research and Students, SIAM, Philadephia, USA, 2002.
\newblock \href {http://dx.doi.org/10.1115/1.1579454} {\path{doi:10.1115/1.1579454}}.

\bibitem{Matcont}
A.~Dhooge, W.~Govaerts, Y.~Kuznetsov, H.~Meijer, B.~Sautois, New features of the software matcont for bifurcation analysis of dynamical systems, Mathematical and Computer Modelling of Dynamical Systems 14~(2) (2008) 147--175.
\newblock \href {http://dx.doi.org/10.1080/13873950701742754} {\path{doi:10.1080/13873950701742754}}.

\bibitem{MATLAB}
The Mathworks, Inc., Natick, Massachusetts, United States, \href{https://www.mathworks.com}{{MATLAB version: 9.13.0 (R2022b)}} (2022).
\newline\urlprefix\url{https://www.mathworks.com}

\bibitem{Julia}
J.~Bezanson, S.~Karpinski, V.~B. Shah, A.~Edelman, Julia: A fast dynamic language for technical computing, arXiv preprint arXiv:1209.5145.

\bibitem{PazoMontbrio2016}
D.~Paz\'o, E.~Montbri\'o, From quasiperiodic partial synchronization to collective chaos in populations of inhibitory neurons with delay, Physical Review Letters 116 (2016) 238101.
\newblock \href {http://dx.doi.org/10.1103/PhysRevLett.116.238101} {\path{doi:10.1103/PhysRevLett.116.238101}}.

\bibitem{DevalleMontbrio2018_delay}
F.~Devalle, E.~Montbri\'o, D.~Paz\'o, Dynamics of a large system of spiking neurons with synaptic delay, Physical Review E 98 (2018) 042214.
\newblock \href {http://dx.doi.org/10.1103/PhysRevE.98.042214} {\path{doi:10.1103/PhysRevE.98.042214}}.

\bibitem{Izhikevich2007}
E.~M. Izhikevich, Dynamical systems in neuroscience : the geometry of excitability and bursting, Computational neuroscience, MIT Press, Cambridge, Mass, 2007.
\newblock \href {http://dx.doi.org/10.7551/mitpress/2526.001.0001} {\path{doi:10.7551/mitpress/2526.001.0001}}.

\bibitem{ermentrout1986parabolic}
G.~B. Ermentrout, N.~Kopell, Parabolic bursting in an excitable system coupled with a slow oscillation, SIAM journal on applied mathematics 46~(2) (1986) 233--253.
\newblock \href {http://dx.doi.org/10.1137/0146017} {\path{doi:10.1137/0146017}}.

\bibitem{Gutkin2022}
B.~Gutkin, Theta neuron model, in: D.~Jaeger, R.~Jung (Eds.), Encyclopedia of Computational Neuroscience, Springer New York, 2022, pp. 3412--3419.
\newblock \href {http://dx.doi.org/10.1007/978-1-0716-1006-0_153} {\path{doi:10.1007/978-1-0716-1006-0_153}}.

\bibitem{Laing2018}
C.~R. Laing, The dynamics of networks of identical theta neurons, Journal of Mathematical Neuroscience 8~(1) (2018) 1--24.
\newblock \href {http://dx.doi.org/10.1186/s13408-018-0059-7} {\path{doi:10.1186/s13408-018-0059-7}}.

\bibitem{Ermentrout2010book}
G.~B. Ermentrout, D.~H. Terman, Mathematical Foundations of Neuroscience, Vol.~35 of Interdisciplinary Applied Mathematics, Springer, New York, 2010.
\newblock \href {http://dx.doi.org/10.1007/978-0-387-87708-2} {\path{doi:10.1007/978-0-387-87708-2}}.

\bibitem{Nicola2013bif}
W.~Nicola, S.~A. Campbell, Bifurcations of large networks of two-dimensional integrate and fire neurons, Journal of Computational Neuroscience 35~(1) (2013) 87--108.
\newblock \href {http://dx.doi.org/10.1007/s10827-013-0442-z} {\path{doi:10.1007/s10827-013-0442-z}}.

\bibitem{ChenCampbell2022_mf}
L.~Chen, S.~A. Campbell, Exact mean‑field models for spiking neural networks with adaptation, Journal of Computational Neuroscience 50~(5) (2022) 445--469.
\newblock \href {http://dx.doi.org/10.1007/s10827-022-00825-9} {\path{doi:10.1007/s10827-022-00825-9}}.

\bibitem{LeeOtt2009_delay}
W.~S. Lee, E.~Ott, T.~M. Antonsen, \href{https://link.aps.org/doi/10.1103/PhysRevLett.103.044101}{Large coupled oscillator systems with heterogeneous interaction delays}, Physical Review Letters 103~(4) (2009) 044101, publisher: American Physical Society.
\newblock \href {http://dx.doi.org/10.1103/PhysRevLett.103.044101} {\path{doi:10.1103/PhysRevLett.103.044101}}.
\newline\urlprefix\url{https://link.aps.org/doi/10.1103/PhysRevLett.103.044101}

\bibitem{WS1994}
S.~Watanabe, S.~H. Strogatz, Constants of motion for superconducting josephson arrays, Physica D: Nonlinear Phenomena 74~(3) (1994) 197--253.
\newblock \href {http://dx.doi.org/10.1016/0167-2789(94)90196-1} {\path{doi:10.1016/0167-2789(94)90196-1}}.

\bibitem{Pazo2016}
D.~Paz{\'o}, E.~Montbri{\'o}, From quasiperiodic partial synchronization to collective chaos in populations of inhibitory neurons with delay, Physical Review Letters 116~(23) (2016) 238101--238101.
\newblock \href {http://dx.doi.org/10.1103/PhysRevLett.116.238101} {\path{doi:10.1103/PhysRevLett.116.238101}}.

\bibitem{Devalle2018_delay}
F.~Devalle, E.~Montbri\'o, D.~Paz\'o, Dynamics of a large system of spiking neurons with synaptic delay, Physical Review E 98 (2018) 042214.
\newblock \href {http://dx.doi.org/10.1103/PhysRevE.98.042214} {\path{doi:10.1103/PhysRevE.98.042214}}.

\bibitem{maple}
M.~B. Monagan, K.~O. Geddes, K.~M. Heal, G.~Labahn, S.~M. Vorkoetter, J.~Devitt, M.~Hansen, D.~Redfern, K.~Rickard, et~al., Maple V Programming Guide: For Release 5, Springer Science \& Business Media, 2012.

\bibitem{Dur-e-Ahmad2012}
M.~Dur{-}e{-}Ahmad, W.~Nicola, S.~A. Campbell, F.~K. Skinner, Network bursting using experimentally constrained single compartment {CA3} hippocampal neuron models with adaptation, Journal of Computational Neuroscience 33~(1) (2012) 21--40.
\newblock \href {http://dx.doi.org/10.1007/s10827-011-0372-6} {\path{doi:10.1007/s10827-011-0372-6}}.

\bibitem{Hemond2008}
P.~Hemond, D.~Epstein, A.~Boley, M.~Migliore, G.~A. Ascoli, D.~B. Jaffe, Distinct classes of pyramidal cells exhibit mutually exclusive firing patterns in hippocampal area ca3b, Hippocampus 18~(4) (2008) 411--424.
\newblock \href {http://dx.doi.org/10.1002/hipo.20404} {\path{doi:10.1002/hipo.20404}}.

\bibitem{Rich2020}
S.~Rich, H.~M. Chameh, M.~Rafiee, K.~Ferguson, F.~K. Skinner, T.~A. Valiante, Inhibitory network bistability explains increased interneuronal activity prior to seizure onset, Frontiers in Neural Circuits 13 (2020) 81.
\newblock \href {http://dx.doi.org/10.3389/fncir.2019.00081} {\path{doi:10.3389/fncir.2019.00081}}.

\bibitem{Nicola2013hetero}
W.~Nicola, S.~A. Campbell, Mean-field models for heterogeneous networks of two-dimensional integrate and fire neurons, Frontiers in Computational Neuroscience 7 (2013) 184.
\newblock \href {http://dx.doi.org/10.3389/fncom.2013.00184} {\path{doi:10.3389/fncom.2013.00184}}.

\bibitem{Buszaki2006}
G.~Buszaki, Rhythms of the Brain, Oxford University Press, Oxford, UK, 2006.
\newblock \href {http://dx.doi.org/10.1093/acprof:oso/9780195301069.001.0001} {\path{doi:10.1093/acprof:oso/9780195301069.001.0001}}.

\bibitem{Vladimirski2008}
B.~B. Vladimirski, J.~Tabak, M.~J. O’Donovan, Episodic activity in a heterogeneous excitatory network, from spiking neurons to mean field, Journal of Computational Neuroscience 25 (2008) 39--63.
\newblock \href {http://dx.doi.org/10.1007/s10827-007-0064-4} {\path{doi:10.1007/s10827-007-0064-4}}.

\bibitem{Kuznetsov_BifBook}
Y.~A. Kuznetsov, Elements of Applied Bifurcation Theory (4th Ed.), Springer, 2023.
\newblock \href {http://dx.doi.org/10.1007/978-3-031-22007-4} {\path{doi:10.1007/978-3-031-22007-4}}.

\bibitem{Bertram2017}
R.~Bertram, J.~E. Rubin, Multi-timescale systems and fast-slow analysis, Mathematical Biosciences 287 (2017) 105--121, 50th Anniversary Issue.
\newblock \href {http://dx.doi.org/10.1016/j.mbs.2016.07.003} {\path{doi:10.1016/j.mbs.2016.07.003}}.

\bibitem{leblanc1996classification}
V.~G. LeBlanc, W.~F. Langford, Classification and unfoldings of 1: 2 resonant hopf bifurcation, Archive for rational mechanics and analysis 136 (1996) 305--357.
\newblock \href {http://dx.doi.org/10.1007/BF02206623} {\path{doi:10.1007/BF02206623}}.

\bibitem{DevalleMontbrio2017}
F.~Devalle, A.~Roxin, E.~Montbri{\'o}, Firing rate equations require a spike synchrony mechanism to correctly describe fast oscillations in inhibitory networks, {PLOS} Computational Biology 13~(12) (2017) e1005881--e1005881.
\newblock \href {http://dx.doi.org/10.1371/journal.pcbi.1005881} {\path{doi:10.1371/journal.pcbi.1005881}}.

\bibitem{lisman2008_CFC}
J.~Lisman, G.~Buzsáki, A neural coding scheme formed by the combined function of gamma and theta oscillations, Schizophrenia Bulletin 34~(5) (2008) 974--980.
\newblock \href {http://dx.doi.org/10.1093/schbul/sbn060} {\path{doi:10.1093/schbul/sbn060}}.

\bibitem{Belluscio2012}
M.~A. Belluscio, K.~Mizuseki, R.~Schmidt, R.~Kempter, G.~Buzs{\'a}ki, Cross-frequency phase{\textendash}phase coupling between theta and gamma oscillations in the hippocampus, Journal of Neuroscience 32~(2) (2012) 423--435.
\newblock \href {http://dx.doi.org/10.1523/JNEUROSCI.4122-11.2012} {\path{doi:10.1523/JNEUROSCI.4122-11.2012}}.

\bibitem{Butler2016}
J.~L. Butler, P.~R.~F. Mendon{\c c}a, H.~P.~C. Robinson, O.~Paulsen, Intrinsic cornu ammonis area 1 theta-nested gamma oscillations induced by optogenetic theta frequency stimulation, Journal of Neuroscience 36~(15) (2016) 4155--4169.
\newblock \href {http://dx.doi.org/10.1523/JNEUROSCI.3150-15.2016} {\path{doi:10.1523/JNEUROSCI.3150-15.2016}}.

\bibitem{Buszaki2012}
G.~Buzs\'{a}ki, X.-J. Wang, Mechanisms of gamma oscillations, Annual Review of Neuroscience 35~(1) (2012) 203--225.
\newblock \href {http://dx.doi.org/10.1146/annurev-neuro-062111-150444} {\path{doi:10.1146/annurev-neuro-062111-150444}}.

\bibitem{Colgin2009_nature}
L.~L. Colgin, T.~Denninger, M.~Fyhn, T.~Hafting, T.~Bonnevie, O.~Jensen, M.-B. Moser, E.~I. Moser, Frequency of gamma oscillations routes flow of information in the hippocampus, Nature 462 (2009) 353–357.
\newblock \href {http://dx.doi.org/10.1038/nature08573} {\path{doi:10.1038/nature08573}}.

\bibitem{canolty2010_CFC}
R.~T. Canolty, R.~T. Knight, The functional role of cross-frequency coupling, Trends in Cognitive Sciences 14~(11) (2010) 506--515.
\newblock \href {http://dx.doi.org/10.1016/j.tics.2010.09.001} {\path{doi:10.1016/j.tics.2010.09.001}}.

\bibitem{aru2015_CFC}
J.~Aru, J.~Aru, V.~Priesemann, M.~Wibral, L.~Lana, G.~Pipa, W.~Singer, R.~Vicente, Untangling cross-frequency coupling in neuroscience, Current Opinion in Neurobiology 31 (2015) 51--61.
\newblock \href {http://dx.doi.org/10.1016/j.conb.2014.08.002} {\path{doi:10.1016/j.conb.2014.08.002}}.

\bibitem{ceni2020_CFC}
A.~Ceni, S.~Olmi, A.~Torcini, D.~Angulo-Garcia, Cross frequency coupling in next generation inhibitory neural mass models, Chaos: An Interdisciplinary Journal of Nonlinear Science 30~(5) (2020) 053121.
\newblock \href {http://dx.doi.org/10.1063/1.5125216} {\path{doi:10.1063/1.5125216}}.

\bibitem{segneri2020_CFC}
M.~Segneri, H.~Bi, S.~Olmi, A.~Torcini, Theta-nested gamma oscillations in next generation neural mass models, Frontiers in Computational Neuroscience 14 (2020) 47.
\newblock \href {http://dx.doi.org/10.3389/fncom.2020.00047} {\path{doi:10.3389/fncom.2020.00047}}.

\end{thebibliography}


\end{document}